\documentclass[10pt,conference]{IEEEtran}
\IEEEoverridecommandlockouts
\usepackage{cite}
\usepackage{amsmath,amssymb}
\usepackage{graphicx}
\usepackage{float}
\usepackage{xcolor}
\usepackage{booktabs}
\usepackage{multirow}
\usepackage{url}
\usepackage{tcolorbox}
\usepackage{microtype}
\usepackage{xspace}
\usepackage{enumitem}
\usepackage{pifont}
\usepackage{fontawesome5}
\usepackage{tikz}
\usepackage{pgfplots}
\usepackage[colorlinks=true,linkcolor=black,citecolor=black,urlcolor=black]{hyperref}
\pgfplotsset{compat=1.18}
\usepgfplotslibrary{fillbetween}
\usetikzlibrary{arrows.meta,positioning,shapes.geometric,shapes.misc,fit,calc,backgrounds}

\AtBeginDocument{%
  }

\title{A Single Patch Is Not Enough: Deterministic Fusion of Repair Candidates}

\author{
\IEEEauthorblockN{
Boyang Yang\textsuperscript{1},
Xiangliang Hu\textsuperscript{2,*},
Luyao Ren\textsuperscript{3,*},
Yanjun Chen\textsuperscript{4}\\
Bach Le\textsuperscript{5},
Tegawend{\'e} F. Bissyand{\'e}\textsuperscript{6},
Haoye Tian\textsuperscript{7,\dag}}
\IEEEauthorblockA{
\textsuperscript{1}Yanshan University;
\textsuperscript{2}Beijing Guoyan Network Data Technology Co., Ltd.}
\IEEEauthorblockA{
\textsuperscript{3}Peking University;
\textsuperscript{4}Google;
\textsuperscript{5}University of Melbourne}
\IEEEauthorblockA{
\textsuperscript{6}University of Luxembourg;
\textsuperscript{7}Aalto University}
\IEEEauthorblockA{
\texttt{yby@ieee.org},
\texttt{huxiangliang@126.com},
\texttt{rly@pku.edu.cn}}
\IEEEauthorblockA{
\texttt{yanjunch98@gmail.com},
\texttt{bach.le@unimelb.edu.au},
\texttt{tegawende.bissyande@uni.lu}}
\IEEEauthorblockA{
\texttt{tianhaoyemail@gmail.com}}
\IEEEauthorblockA{\textsuperscript{*}Equal contribution. \quad \textsuperscript{\dag}Corresponding author.}
}

\newcommand{\method}{PatchFusion\xspace}
\newcommand{\dataset}{PatchFuseBench\xspace}

\providecommand{\Description}[1]{}
\newcommand{\find}[1]{
\begin{tcolorbox}[leftrule=1mm,toprule=0mm,bottomrule=0mm,left=1pt,right=2pt,top=2pt,bottom=2pt]
\em #1
\end{tcolorbox}
}
\newcommand{\findx}[1]{\find{#1}}
\newcommand{\findtag}[1]{\textnormal{\bfseries #1}}
\newcommand{\keyquote}[1]{%
\begin{tcolorbox}[colback=black!3!white,colframe=black!70!white,boxrule=0.5pt,left=4pt,right=4pt,top=4pt,bottom=4pt]
\em #1
\end{tcolorbox}}
\newcommand{\gainpos}[1]{\textcolor{teal!65!black}{#1}}
\newcommand{\gainneg}[1]{\textcolor{red!70!black}{#1}}
\newcommand{\scoregain}[2]{#1~{\normalfont(#2)}}

\begin{document}
\bstctlcite{IEEEexample:BSTcontrol}

\maketitle

\begin{abstract}
Modern LLM coding agents are commonly evaluated using pass@$k$, but developers typically apply a single final patch in real-world settings.
This pass@$k$-to-pass@1 gap is a post-generation problem: a candidate patch pool may contain a correct patch, but the system must decide which one to suggest to developers.
Existing post-generation approaches mainly rank whole candidates, filter them with tests, or query an LLM judge, but none deterministically reuse shared edit-atom evidence to both select and construct the final patch.
Thus, we propose \method, a deterministic atomic evidence fusion approach for candidate patches that consults no test outcome at decision time.
\method first fuses whole-diff agreement into a repair neighborhood, selects an auditable representative, and then applies evidence-constrained fusion (ECF) to retain repeated edit atoms and prune unsupported parts.
To evaluate this setting, we build \dataset, a fixed-pool benchmark covering SWE-bench Verified, SWE-bench Multilingual, and Defects4J candidate patches.

On \dataset, \method solves 426/500 bugs on SWE-bench Verified and 236/300 on SWE-bench Multilingual, and reaches 87/371 plausible patches on Defects4J, outperforming every matched candidate-pool selector on all three.
\method recovers 41 and 27 bugs that no single source solves (30 and 18 more over the best single source), repairs that require combining evidence across candidates rather than choosing one patch.
It decides in only 3.28 ms per bug, consulting no test or model at decision time.
Ablation studies show that ECF adds $+5/+6/+9$ solved bugs by recovering in-pool repairs that selection misses, with no observed regression, and that \method's gains remain stable as candidate pools are resampled.
On these complementary multi-source pools, cross-candidate evidence recovers more correct patches than the test-based and LLM-based selectors we evaluate, at orders-of-magnitude lower cost, reaching within 96.2\% and 89.7\% of the candidate-reachable ceiling on the two SWE-bench benchmarks.
\end{abstract}

\section{Introduction}

\begin{figure*}[t]
  \centering
  \scriptsize
  \definecolor{dfBorder}{HTML}{D0D7DE}
  \definecolor{dfHdrBg}{HTML}{F1F2F4}
  \definecolor{dfHdrTx}{HTML}{6E7681}
  \definecolor{dfCode}{HTML}{1F2328}
  \definecolor{dfDelBg}{HTML}{FFEBE9}
  \definecolor{dfDelGut}{HTML}{FFCECB}
  \definecolor{dfDelSn}{HTML}{CF222E}
  \definecolor{dfAddBg}{HTML}{E6FFEC}
  \definecolor{dfAddGut}{HTML}{BBF0C9}
  \definecolor{dfAddSn}{HTML}{1A7F37}
  \newcommand{\onehunk}[8]{%
    \begin{scope}
      \clip[rounded corners=2.6pt] (#1,#2) rectangle (#1+#3,#2-\hh-2*\rh);
      \fill[dfHdrBg] (#1,#2) rectangle (#1+#3,#2-\hh);
      \fill[dfDelBg] (#1,#2-\hh) rectangle (#1+#3,#2-\hh-\rh);
      \fill[dfDelGut] (#1,#2-\hh) rectangle (#1+\gw,#2-\hh-\rh);
      \fill[dfAddBg] (#1,#2-\hh-\rh) rectangle (#1+#3,#2-\hh-2*\rh);
      \fill[dfAddGut] (#1,#2-\hh-\rh) rectangle (#1+\gw,#2-\hh-2*\rh);
      \draw[dfBorder,line width=0.3pt] (#1,#2-\hh) -- (#1+#3,#2-\hh);
      \draw[dfBorder,line width=0.3pt] (#1+\gw,#2-\hh) -- (#1+\gw,#2-\hh-2*\rh);
    \end{scope}
    \node[anchor=base west,text=dfHdrTx,font=\fontsize{6.4}{7.4}\selectfont\ttfamily] at (#1+\tp,#2-\hh+0.155) {#4};
    \node[anchor=base east,text=#6,font=\sffamily\bfseries\scriptsize] at (#1+#3-0.14,#2-\hh+0.165) {#5};
    \node[anchor=base,text=dfDelSn,font=\ttfamily\scriptsize] at (#1+\gw/2,#2-\hh-\rh+0.145) {-};
    \node[anchor=base west,text=dfCode,font=\ttfamily\scriptsize] at (#1+\gw+\tp,#2-\hh-\rh+0.145) {#7};
    \node[anchor=base,text=dfAddSn,font=\ttfamily\scriptsize] at (#1+\gw/2,#2-\hh-2*\rh+0.145) {+};
    \node[anchor=base west,text=dfCode,font=\ttfamily\scriptsize] at (#1+\gw+\tp,#2-\hh-2*\rh+0.145) {#8};
    \draw[dfBorder,line width=0.5pt,rounded corners=2.6pt] (#1,#2) rectangle (#1+#3,#2-\hh-2*\rh);
  }
  \resizebox{\textwidth}{!}{%
  \begin{tikzpicture}[
    font=\sffamily\scriptsize,
    txt/.style={align=left, text width=42mm, inner sep=0pt},
    head/.style={txt, font=\sffamily\bfseries\footnotesize},
    panelframe/.style={rounded corners=5pt, line width=0.72pt},
    tag/.style={
      draw, rounded corners=2pt, line width=0.45pt, fill=white,
      inner xsep=2.6pt, inner ysep=1.6pt, font=\sffamily\bfseries\scriptsize
    },
    arr/.style={-{Stealth[length=2.0mm]}, line width=0.72pt, draw=black!58, shorten >=1.4pt, shorten <=1.4pt}
  ]
    \pgfdeclarelayer{panels}
    \pgfsetlayers{panels,main}

    \def\colpitch{5.35}
    \def\paneltop{0.30}
    \def\panelbot{-4.30}
    \def\gw{0.34}\def\hh{0.46}\def\rh{0.42}\def\tp{0.14}\def\bw{4.24}

    \begin{pgfonlayer}{panels}
      \draw[panelframe, fill=blue!5,   draw=blue!58!black]
        (-0.28,-0.75) rectangle (4.52,-3.25);
      \draw[panelframe, fill=red!5,    draw=red!55!black]
        (\colpitch-0.28,\paneltop) rectangle (\colpitch+4.52,\panelbot);
      \draw[panelframe, fill=purple!5, draw=purple!58!black]
        (2*\colpitch-0.28,\paneltop) rectangle (2*\colpitch+4.52,\panelbot);
      \draw[panelframe, fill=green!7,  draw=green!55!black]
        (3*\colpitch-0.28,-0.50) rectangle (3*\colpitch+4.52,-3.50);
    \end{pgfonlayer}

    \node[head, anchor=north west] at (0,-1.09)
      {Empty \texttt{env} clears the environment};
    \begin{scope}
      \clip[rounded corners=2.6pt] (0,-1.78) rectangle (\bw,-1.78-\hh-2*\rh);
      \fill[dfHdrBg] (0,-1.78) rectangle (\bw,-1.78-\hh);
      \fill[white] (0,-1.78-\hh) rectangle (\bw,-1.78-\hh-2*\rh);
      \draw[dfBorder,line width=0.3pt] (0,-1.78-\hh) -- (\bw,-1.78-\hh);
    \end{scope}
    \node[anchor=base west,text=dfHdrTx,font=\fontsize{6.4}{7.4}\selectfont\ttfamily] at (\tp,-1.78-\hh+0.155) {postgresql/client.py};
    \node[anchor=base west,text=dfCode,font=\ttfamily\scriptsize] at (\tp,-1.78-\hh-\rh+0.145)
      {return args, \textcolor{dfDelSn}{\bfseries env}};
    \node[anchor=base west,text=dfCode,font=\ttfamily\scriptsize] at (\tp,-1.78-\hh-2*\rh+0.145)
      {subprocess.run(.., \textcolor{dfDelSn}{\bfseries env=env})};
    \draw[dfBorder,line width=0.5pt,rounded corners=2.6pt] (0,-1.78) rectangle (\bw,-1.78-\hh-2*\rh);

    \node[head, anchor=north west] at (\colpitch,-0.12)
      {Each candidate fixes only one side};
    \onehunk{\colpitch}{-0.95}{\bw}{producer fix}{unresolved}{dfDelSn}
      {return args, env}{return args, env or None}
    \onehunk{\colpitch}{-2.50}{\bw}{consumer fix}{unresolved}{dfDelSn}
      {run(.., env=env)}{run(.., env=env or None)}
    \node[txt, anchor=south west, text=red!60!black] at (\colpitch,\panelbot+0.16)
      {$\times$~\textbf{0 / 6 candidates resolved}};

    \node[head, anchor=north west] at (2*\colpitch,-0.12)
      {Edits repeated across candidates};
    \onehunk{2*\colpitch}{-0.95}{\bw}{settings\_to\_cmd\_args\_env}{$\times 3$}{dfAddSn}
      {return args, env}{return args, env or None}
    \onehunk{2*\colpitch}{-2.50}{\bw}{runshell}{$\times 3$}{dfAddSn}
      {run(.., env=env)}{run(.., env=env or None)}

    \node[head, anchor=north west] at (3*\colpitch,-0.83)
      {Not a submitted candidate};
    \def\fx{3*\colpitch}\def\fy{-1.20}
    \begin{scope}
      \clip[rounded corners=2.6pt] (\fx,\fy) rectangle (\fx+\bw,\fy-\hh-4*\rh);
      \fill[dfHdrBg] (\fx,\fy) rectangle (\fx+\bw,\fy-\hh);
      \fill[dfDelBg] (\fx,\fy-\hh) rectangle (\fx+\bw,\fy-\hh-\rh);
      \fill[dfDelGut] (\fx,\fy-\hh) rectangle (\fx+\gw,\fy-\hh-\rh);
      \fill[dfAddBg] (\fx,\fy-\hh-\rh) rectangle (\fx+\bw,\fy-\hh-2*\rh);
      \fill[dfAddGut] (\fx,\fy-\hh-\rh) rectangle (\fx+\gw,\fy-\hh-2*\rh);
      \fill[dfDelBg] (\fx,\fy-\hh-2*\rh) rectangle (\fx+\bw,\fy-\hh-3*\rh);
      \fill[dfDelGut] (\fx,\fy-\hh-2*\rh) rectangle (\fx+\gw,\fy-\hh-3*\rh);
      \fill[dfAddBg] (\fx,\fy-\hh-3*\rh) rectangle (\fx+\bw,\fy-\hh-4*\rh);
      \fill[dfAddGut] (\fx,\fy-\hh-3*\rh) rectangle (\fx+\gw,\fy-\hh-4*\rh);
      \draw[dfBorder,line width=0.3pt] (\fx,\fy-\hh) -- (\fx+\bw,\fy-\hh);
      \draw[dfBorder,line width=0.3pt] (\fx,\fy-\hh-2*\rh) -- (\fx+\bw,\fy-\hh-2*\rh);
      \draw[dfBorder,line width=0.3pt] (\fx+\gw,\fy-\hh) -- (\fx+\gw,\fy-\hh-4*\rh);
    \end{scope}
    \node[anchor=base west,text=dfHdrTx,font=\fontsize{6.4}{7.4}\selectfont\ttfamily] at (\fx+\tp,\fy-\hh+0.155) {fused patch};
    \node[anchor=base east,text=dfAddSn,font=\sffamily\bfseries\scriptsize] at (\fx+\bw-0.14,\fy-\hh+0.165) {\checkmark~resolved};
    \node[anchor=base,text=dfDelSn,font=\ttfamily\scriptsize] at (\fx+\gw/2,\fy-\hh-\rh+0.145) {-};
    \node[anchor=base west,text=dfCode,font=\ttfamily\scriptsize] at (\fx+\gw+\tp,\fy-\hh-\rh+0.145) {return args, env};
    \node[anchor=base,text=dfAddSn,font=\ttfamily\scriptsize] at (\fx+\gw/2,\fy-\hh-2*\rh+0.145) {+};
    \node[anchor=base west,text=dfCode,font=\ttfamily\scriptsize] at (\fx+\gw+\tp,\fy-\hh-2*\rh+0.145) {return args, env or None};
    \node[anchor=base,text=dfDelSn,font=\ttfamily\scriptsize] at (\fx+\gw/2,\fy-\hh-3*\rh+0.145) {-};
    \node[anchor=base west,text=dfCode,font=\ttfamily\scriptsize] at (\fx+\gw+\tp,\fy-\hh-3*\rh+0.145) {run(.., env=env)};
    \node[anchor=base,text=dfAddSn,font=\ttfamily\scriptsize] at (\fx+\gw/2,\fy-\hh-4*\rh+0.145) {+};
    \node[anchor=base west,text=dfCode,font=\ttfamily\scriptsize] at (\fx+\gw+\tp,\fy-\hh-4*\rh+0.145) {run(.., env=env or None)};
    \draw[dfBorder,line width=0.5pt,rounded corners=2.6pt] (\fx,\fy) rectangle (\fx+\bw,\fy-\hh-4*\rh);

    \node[tag, text=blue!58!black,   draw=blue!58!black]   at (2.12,-0.75) {buggy behavior};
    \node[tag, text=red!55!black,    draw=red!55!black]    at (\colpitch+2.12,\paneltop) {failed candidates};
    \node[tag, text=purple!58!black, draw=purple!58!black] at (2*\colpitch+2.12,\paneltop) {ECF fusion};
    \node[tag, text=green!45!black,  draw=green!45!black]  at (3*\colpitch+2.12,-0.50) {fused patch};

    \def\arrowy{-2.00}
    \draw[arr] (4.61,\arrowy) -- (\colpitch-0.35,\arrowy);
    \draw[arr] (\colpitch+4.61,\arrowy) -- (2*\colpitch-0.35,\arrowy);
    \draw[arr] (2*\colpitch+4.61,\arrowy) -- (3*\colpitch-0.35,\arrowy);
  \end{tikzpicture}%
  }
  \caption{Motivating example (\texttt{django\_\_django-14315}), where \method fuses repeated edit atoms into a resolved patch no candidate submitted.}
  \Description{A four-panel motivating figure based on SWE-bench Verified django\_\_django-14315. Panel 1: an empty env dictionary clears the inherited environment in the producer return value and the consumer subprocess call. Panel 2: two failed candidates, one fixing only the producer side and one only the consumer side, both unresolved; none of the six candidates is resolved. Panel 3: the two correct edits, in the settings_to_cmd_args_env and runshell scopes, each repeated across three candidates. Panel 4: a fused repository-backed patch, not a submitted candidate, resolved by the official runner.}
  \label{fig:motivating-example}
\end{figure*}
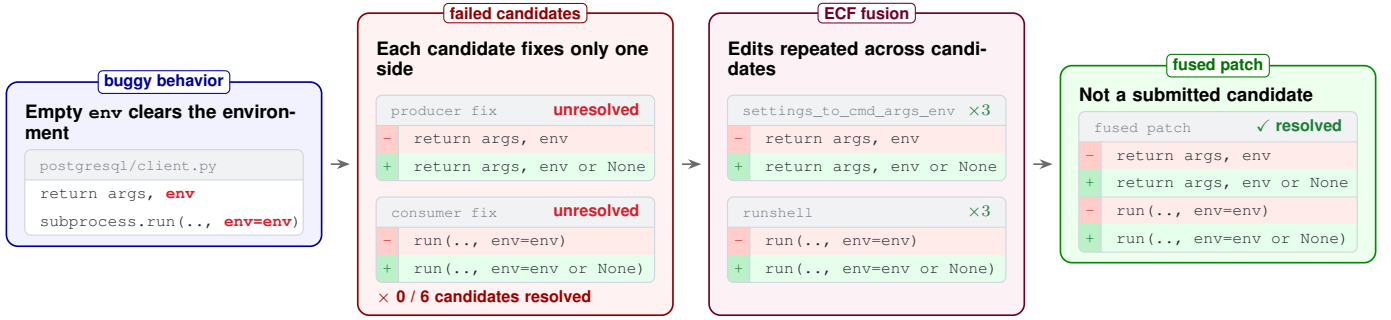

LLM coding agents have improved rapidly and now resolve a large and growing share of real-world software issues, from bug fixes to repository-level edits~\cite{fan2023aprllm,xia2023llmapr,bouzenia2025repairagent}.
However, behind each reported result, there is not one answer but many: a coding agent samples $k$ candidate patches per bug and is scored by pass@$k$, which counts a bug as fixed when any of the $k$ samples is correct~\cite{yang2024morepair}.
A developer applies only a single pass@1 patch, because validating and reviewing every candidate is budgeted work~\cite{beller2014modernreview,mcintosh2016reviewquality} that prior repair-targeted studies also quantify~\cite{huq2022review4repair,kirbas2021bloomberg}. Because pass@$k$ can far exceed pass@1, repairs that already sit in the sampled pool are lost the moment one patch must be chosen~\cite{yang2024cref}.
This loss grows once the pool spans systems: on SWE-bench~\cite{jimenez2024swebench}, different agents and models fix overlapping but distinct bugs, and combining their patches covers far more than the best single system~\cite{li2022alphacode,ehrlich2025codemonkeys,antoniades2024swesearch}, reaching 443 of 500 against 396 on Verified and 263 of 300 against 218 across 7 models of a multilingual pool.
A fixed candidate pool therefore already holds many more repairs than the one patch a developer applies; the open problem is how to surface them.

\keyquote{\textbf{Correct fix may already be in the attempts. It just has to be pieced together from the parts each of them got right.}}

Surfacing them is a post-generation problem: once the pool is fixed, the system must commit to a single patch without re-running the sampling that produced it.
This is hard because a candidate that looks plausible is often still wrong~\cite{qi2015plausibility,petke2024patchplausibility}, while the correct fix is usually present in the pool yet easily mis-selected, and the edits that compose it often recur across several otherwise-imperfect candidates.
The common response is to select or rank whole candidates by how far they agree~\cite{yang2025llmsurvey}, which is cheap enough for a pool whose expensive sampling is already paid for; methods that recover more typically do so by spending test runs or extra model calls per bug.

However, existing approaches still face \emph{two major limitations}:

\begin{itemize}[leftmargin=*]
\item[\ding{172}] \textbf{Patch selectors are limited to the best whole patch.}
Millisecond-cost selectors treat the pool as a set of whole-diff alternatives and score each submitted diff as one unit: Agentless keeps the patch that the most candidates reproduce exactly~\cite{xia2025agentless}, medoid and similarity selectors return the diff most central to the pool~\cite{ghanbari2022similarityranking,li2022alphacode}, and rank-aggregation selectors such as Borda, Copeland, and MC4 merge several orderings into one~\cite{dwork2001rankaggregation}.
All of them credit a fix only when an entire candidate already matches, so they cannot exceed the per-bug candidate-reachable ceiling and cannot recover a fix that no single candidate fully expresses; when no whole candidate is correct, the repair already present in fragments is lost.
The correct local edits often recur across several otherwise-failing candidates, evidence that is free because it comes from the diffs already available, yet whole-diff scoring discards it.
\item[\ding{173}] \textbf{Effective methods pay for each bug in tests or model calls.}
The selectors that beat simple consensus buy their gains per bug: test and dynamic-impact selectors re-run the suite or measure execution differences~\cite{ghanbari2022impact,tian2022failingtests}, learned correctness predictors score each patch with a trained model~\cite{xiong2018patchcorrectness,ye2021patchassessment}, LLM-as-judge selection prompts a model to rank the candidates~\cite{zhou2024llmapca}, and generative fusion such as self-consistency, CodeT, and universal self-consistency asks a model to synthesize a new fused program~\cite{wang2023selfconsistency,chen2023codet,chen2024usc}.
Each adds seconds of inference, a dollar cost, and a dependence on an online model at the step that should add only milliseconds, the marginal decision over a pool that was already paid for.
Beyond this cost, generative fusion carries a further risk: the program it emits need not trace to any candidate and is unverified before validation, which is less auditable when the budget allows only one patch.
\end{itemize}

\textbf{This paper.}
To close these gaps, we propose \method, a deterministic patch-\emph{fusion} approach for fixed candidate pools.
Unlike free-form generative fusion, \method recombines only repeated edit evidence and keeps every output auditable against the repository state or the submitted candidates, all at \emph{decision time}, before tests are run or run-quality metadata is available.
Deciding without test outcomes is not only cheaper but harder to exploit: passing the provided tests does not imply a correct fix~\cite{qi2015plausibility,petke2024patchplausibility}, and as repair models increasingly optimize against the evaluation tests, any selector that trusts test outcomes inherits that exploit, whereas \method's static cross-candidate agreement never consults a test, so it does not add another test-outcome-dependent selection channel.
\method completes fusion in four steps: it (i) collapses exact duplicates into support counts, (ii) fuses whole-diff agreement into a repair neighborhood, (iii) selects an auditable representative under the pool structure, and (iv) applies evidence-constrained fusion (ECF) to keep edit atoms repeated across candidates and prune unsupported ones.

On SWE-bench Verified, SWE-bench Multilingual, and Defects4J, \method solves 426/500 and 236/300 SWE-bench bugs and produces 87/371 plausible Defects4J patches, repairing 38.0 and 23.7 more bugs than uniform random selection and outperforming consensus and LLM-based selectors.
Ablation studies show that ECF adds $+5/+6/+9$ solved bugs across the three benchmarks with no observed regression.
\method runs in 3.28 ms per bug, two to three orders of magnitude below the model-based selectors.
These results indicate that static cross-candidate agreement, rather than test-based validation or model judgments, can expose most reachable repairs at low cost.

The main contributions of our work are as follows:

\begin{itemize}[noitemsep,topsep=0pt,leftmargin=2em]
\item {\bf Approach.}
We propose \method, a deterministic four-step patch-fusion approach that combines exact support, repair-neighborhood fusion, representative selection, and evidence-constrained fusion (ECF) to recombine repeated edit evidence into an auditable patch.

\item {\bf Benchmark.}
We construct \dataset, a reusable patch-fusion benchmark that packages fixed multi-system candidate pools from public leaderboard and repair runs on SWE-bench Verified, SWE-bench Multilingual, and Defects4J together with cached official labels, so any pooled candidate is scored without rerunning tests, while newly generated patches are scored by the official oracle.

\item {\bf Experiments.}
On all three pools, \method solves more bugs than uniform random selection, external ranking/consensus baselines, and strong DeepSeek-V4-Pro baselines.

\item {\bf Analysis.}
An ablation and a case study show that ECF is a content-fusion layer that recovers mis-selected repairs with no observed regression, constructing the submitted patch from repeated atoms (absent from the pool in 8 SWE-bench cases) rather than copying a candidate.

\end{itemize}

\section{Motivating Example}\label{sec:motivating-example}

Figure~\ref{fig:motivating-example} shows a concrete case from SWE-bench Verified, \texttt{django\_\_django-14315}.
The bug is in Django's \texttt{dbshell}: when launching the database shell, the client passes \texttt{env=\{\}} to \texttt{subprocess.run}, which replaces the child environment with an empty one instead of inheriting \texttt{os.environ}.
The fix is to pass \texttt{env or None} so the child inherits the parent environment, and it spans 2 scopes: the PostgreSQL client must return \texttt{env or None} from \texttt{settings\_to\_cmd\_args\_env}, and the base client must forward \texttt{env or None} into \texttt{subprocess.run} in \texttt{runshell}.

In this candidate pool, \emph{no submitted candidate solves the bug}: all 6 candidates are unresolved by the official runner.
The failures are complementary rather than redundant.
Some candidates repair only the producer side (the returned environment), others repair only the consumer side (the \texttt{subprocess.run} call), and the rest add tests or comments without completing the fix.
A ranking or selection method can only return one of these whole diffs unchanged, so it cannot solve the bug: the correct repair exists in the pool only as separated fragments, never as a complete patch.

Evidence-constrained fusion (ECF) instead reasons below the whole-diff level.
It decomposes candidates into scope-local edit atoms and keeps those repeated across candidates (the producer-side atom \texttt{return args, env or None} and the consumer-side atom \texttt{env=env or None}, each supported by 3 candidates), while dropping unsupported extras such as added tests or comments.
Fusing these 2 supported atoms across both scopes reconstructs a single repository-backed patch that the official SWE-bench runner resolves.

This patch is \emph{not} an exact submitted candidate.
The pool contained 0 solved patches, yet \method assembles a solved one from the locally correct edits of several failed candidates.
This is what separates patch-level evidence fusion from candidate selection: \method recovers a repair that no single candidate expresses, while keeping the output auditable, since it is reconstructed against the checked-out repository state and traced back to the supported candidate atoms.
The same audit rule holds throughout our evaluation; Defects4J pass@10 pools expose method bodies for atom-level reconstruction, and SWE-bench pools additionally allow scope reconstruction from the checked-out repository, so ECF can submit a new but auditable diff when the static guards are satisfied.

\section{Approach}
\label{sec:approach}

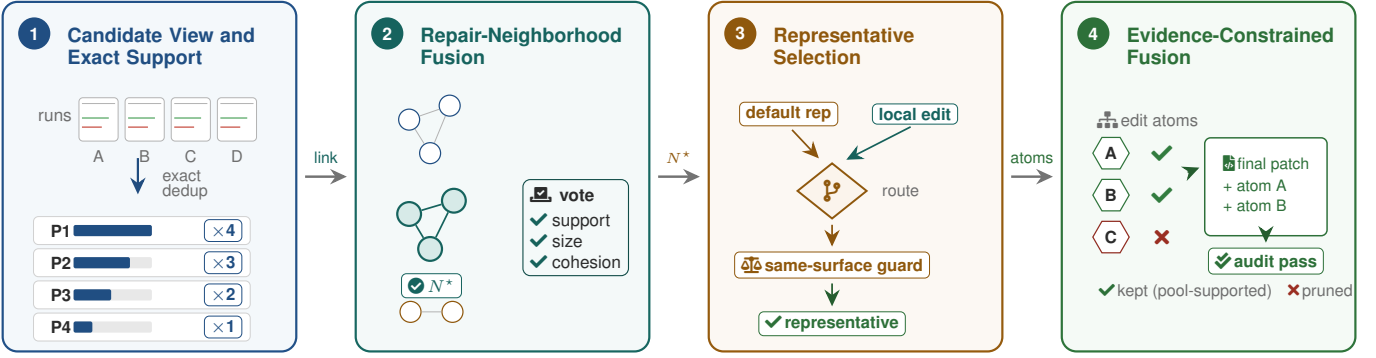
\begin{figure*}[t]
  \centering
  \definecolor{cBlue}{HTML}{2D6CB5}
  \definecolor{cTeal}{HTML}{1F8A82}
  \definecolor{cAmber}{HTML}{C77A12}
  \definecolor{cGreen}{HTML}{3E9B4F}
  \definecolor{cInk}{HTML}{2B2B2B}
  \definecolor{cMute}{HTML}{6E6E6E}
  \definecolor{cKeep}{HTML}{3E9B4F}
  \definecolor{cPrune}{HTML}{C0392B}
  \resizebox{\textwidth}{!}{%
  \begin{tikzpicture}[
    font=\sffamily,
    panel/.style={rounded corners=6pt, line width=0.9pt, draw=#1!75!black,
                  fill=#1!5, minimum width=42mm, minimum height=50mm},
    htitle/.style={font=\sffamily\bfseries\footnotesize, text=#1!72!black, align=left},
    hnum/.style={circle, fill=#1!72!black, text=white, font=\sffamily\bfseries\scriptsize,
                 inner sep=0pt, minimum size=4.6mm},
    kw/.style={font=\sffamily\bfseries\scriptsize, text=cInk},
    klab/.style={font=\sffamily\scriptsize, text=cMute},
    tny/.style={font=\sffamily\fontsize{6.6}{7.6}\selectfont, text=cMute},
    chip/.style={rounded corners=2pt, draw=#1!72!black, line width=0.4pt, fill=white,
                 inner xsep=2.6pt, inner ysep=1.6pt,
                 font=\sffamily\bfseries\fontsize{6.8}{7.6}\selectfont, text=#1!72!black},
    card/.style={rounded corners=1.6pt, draw=black!30, line width=0.4pt, fill=white,
                 minimum width=5.6mm, minimum height=6.2mm, inner sep=0pt},
    prow/.style={rounded corners=2pt, draw=black!18, line width=0.4pt, fill=white,
                 minimum width=31mm, minimum height=4.0mm, inner sep=0pt},
    gnode/.style={circle, draw=#1!72!black, line width=0.5pt, fill=white,
                  minimum size=3.0mm, inner sep=0pt},
    flow/.style={-{Stealth[length=2.4mm,width=2.2mm]}, line width=1.0pt, draw=black!55,
                 shorten >=2pt, shorten <=2pt},
    thin/.style={line width=0.35pt, draw=black!30}
  ]
    \def\pitch{5.05}
    \def\midy{-2.55}

    \foreach \i/\col/\ttl in {%
       0/cBlue/{Candidate View and\protect\\ Exact Support},
       1/cTeal/{Repair-Neighborhood\protect\\ Fusion},
       2/cAmber/{Representative\protect\\ Selection},
       3/cGreen/{Evidence-Constrained\protect\\ Fusion}}{
      \node[panel=\col, anchor=north west] (P\i) at (\i*\pitch,0) {};
      \node[hnum=\col, anchor=north west] at ($(P\i.north west)+(0.30,-0.30)$) {\the\numexpr\i+1\relax};
      \node[htitle=\col, anchor=north west, text width=34mm] at ($(P\i.north west)+(0.82,-0.26)$) {\ttl};
    }

    \node[klab, anchor=north west] at ($(P0.north west)+(0.40,-1.46)$) {runs};
    \foreach \k/\r in {0/A,1/B,2/C,3/D}{
      \node[card, anchor=north west] (rc\k) at ($(P0.north west)+(1.10+\k*0.66,-1.38)$) {};
      \draw[thin] ($(rc\k.north west)+(0.05,-0.09)$)--($(rc\k.north east)+(-0.05,-0.09)$);
      \draw[cKeep, line width=0.45pt] ($(rc\k.west)+(0.05,0.02)$)--($(rc\k.east)+(-0.14,0.02)$);
      \draw[cPrune, line width=0.45pt] ($(rc\k.west)+(0.05,-0.11)$)--($(rc\k.east)+(-0.24,-0.11)$);
      \node[tny, anchor=north] at ($(rc\k.south)+(0,0.0)$) {\r};
    }
    \draw[flow, draw=cBlue!70!black, line width=0.9pt] ($(P0.north west)+(1.95,-2.18)$)--($(P0.north west)+(1.95,-2.84)$);
    \node[tny, anchor=west] at ($(P0.north west)+(2.18,-2.50)$) {exact};
    \node[tny, anchor=west] at ($(P0.north west)+(2.18,-2.74)$) {dedup};
    \foreach \p/\s/\fr/\yy in {1/4/1.00/-3.08, 2/3/0.72/-3.54, 3/2/0.48/-4.00, 4/1/0.24/-4.46}{
      \node[prow, anchor=north west] (pr\p) at ($(P0.north west)+(0.45,\yy)$) {};
      \node[kw, anchor=west, font=\sffamily\bfseries\fontsize{6.8}{7.6}\selectfont] at ($(pr\p.west)+(0.13,0)$) {P\p};
      \fill[black!9, rounded corners=0.8pt] ($(pr\p.west)+(0.58,-0.08)$) rectangle ($(pr\p.west)+(1.70,0.08)$);
      \fill[cBlue!72!black, rounded corners=0.8pt] ($(pr\p.west)+(0.58,-0.08)$) rectangle ($(pr\p.west)+(0.58+1.12*\fr,0.08)$);
      \node[chip=cBlue, anchor=east] at ($(pr\p.east)+(-0.12,0)$) {$\times$\s};
    }

    \node[gnode=cBlue] (a1) at ($(P1.north west)+(0.78,-1.70)$) {};
    \node[gnode=cBlue] (a2) at ($(P1.north west)+(1.36,-1.50)$) {};
    \node[gnode=cBlue] (a3) at ($(P1.north west)+(1.08,-2.18)$) {};
    \draw[thin] (a1)--(a2) (a2)--(a3) (a1)--(a3);
    \node[gnode=cTeal, fill=cTeal!18, line width=0.9pt, minimum size=3.4mm] (b1) at ($(P1.north west)+(0.78,-3.05)$) {};
    \node[gnode=cTeal, fill=cTeal!18, line width=0.9pt, minimum size=3.4mm] (b2) at ($(P1.north west)+(1.42,-2.86)$) {};
    \node[gnode=cTeal, fill=cTeal!18, line width=0.9pt, minimum size=3.4mm] (b3) at ($(P1.north west)+(1.08,-3.55)$) {};
    \draw[cTeal!72!black, line width=0.7pt] (b1)--(b2) (b2)--(b3) (b1)--(b3);
    \begin{scope}[on background layer]
      \node[draw=cTeal!72!black, line width=0.8pt, dash pattern=on 1.4pt off 1.2pt,
            rounded corners=4pt, fit=(b1)(b2)(b3), inner sep=2.2pt, fill=cTeal!7] (nstar) {};
    \end{scope}
    \node[chip=cTeal, anchor=north] at ($(nstar.south)+(0,-0.07)$) {\faCheckCircle\,$N^\star$};
    \node[gnode=cAmber] (c1) at ($(P1.north west)+(0.80,-4.45)$) {};
    \node[gnode=cAmber] (c2) at ($(P1.north west)+(1.40,-4.45)$) {};
    \draw[thin] (c1)--(c2);
    \node[rounded corners=3pt, draw=cTeal!55!black, fill=cTeal!5, line width=0.45pt,
          anchor=north west, minimum width=15mm, align=left, inner xsep=3pt, inner ysep=3pt,
          font=\sffamily\fontsize{7}{8.6}\selectfont, text=cInk]
       (score) at ($(P1.north west)+(2.40,-2.55)$)
       {{\bfseries\faVoteYea\ \,vote}\\[1.5pt]
        \textcolor{cTeal!70!black}{\faCheck}~support\\
        \textcolor{cTeal!70!black}{\faCheck}~size\\
        \textcolor{cTeal!70!black}{\faCheck}~cohesion};

    \node[chip=cAmber, anchor=west] (inD) at ($(P2.north west)+(0.45,-1.62)$) {default rep};
    \node[chip=cTeal, anchor=west] (inL) at ($(P2.north west)+(2.35,-1.62)$) {local edit};
    \node[diamond, draw=cAmber!72!black, line width=0.8pt, fill=cAmber!8,
          inner sep=0pt, minimum width=10mm, minimum height=8mm] (router)
          at ($(P2.north west)+(1.78,-2.72)$) {};
    \node[font=\fontsize{9}{9}\selectfont, text=cAmber!72!black] at (router.center) {\faCodeBranch};
    \node[tny, anchor=west] at ($(router.east)+(0.08,0)$) {route};
    \draw[flow, draw=cAmber!70!black, line width=0.7pt] (inD.south)-- ($(router.north)+(-0.14,0)$);
    \draw[flow, draw=cTeal!70!black, line width=0.7pt] (inL.south)-- ($(router.north)+(0.14,0)$);
    \node[chip=cAmber, anchor=center] (guard) at ($(P2.north west)+(1.78,-3.78)$) {\faBalanceScale\,same-surface guard};
    \draw[flow, draw=cAmber!70!black, line width=0.7pt] (router.south)--(guard.north);
    \node[chip=cGreen, anchor=center, font=\sffamily\bfseries\fontsize{6.8}{7.6}\selectfont]
       (rep) at ($(P2.north west)+(1.78,-4.62)$) {\faCheck\,representative};
    \draw[flow, draw=cGreen!60!black, line width=0.7pt] (guard.south)--(rep.north);

    \node[klab, anchor=north west, font=\sffamily\fontsize{6.8}{7.6}\selectfont] at ($(P3.north west)+(0.40,-1.46)$) {\faSitemap\,edit atoms};
    \foreach \k/\nm/\vc/\vi/\yy in {%
        0/A/cKeep/{\faCheck}/-2.18,
        1/B/cKeep/{\faCheck}/-2.78,
        2/C/cPrune/{\faTimes}/-3.38}{
      \node[regular polygon, regular polygon sides=6, draw=\vc!75!black, line width=0.5pt,
            fill=white, inner sep=0pt, minimum size=5.2mm] (at\k) at ($(P3.north west)+(0.72,\yy)$) {};
      \node[font=\sffamily\bfseries\fontsize{6.8}{7.6}\selectfont, text=cInk] at (at\k.center) {\nm};
      \node[font=\fontsize{8.5}{9}\selectfont, text=\vc!80!black] at ($(P3.north west)+(1.45,\yy)$) {\vi};
    }
    \node[tny, anchor=north west] at ($(P3.north west)+(0.42,-3.92)$)
      {\textcolor{cKeep!72!black}{\faCheck}\,kept (pool-supported)\quad\textcolor{cPrune!80!black}{\faTimes}\,pruned};
    \node[rounded corners=2.4pt, draw=cGreen!72!black, line width=0.6pt, fill=white,
          anchor=north west, minimum width=17.5mm, minimum height=14mm, inner sep=0pt]
          (finaldiff) at ($(P3.north west)+(2.05,-1.95)$) {};
    \node[anchor=north west, font=\sffamily\fontsize{6.4}{7.8}\selectfont, text=cKeep!72!black, align=left]
      at ($(finaldiff.north west)+(0.15,-0.13)$) {\faFileCode\,final patch\\[1.5pt]
       \textcolor{cKeep!72!black}{+ atom A}\\\textcolor{cKeep!72!black}{+ atom B}};
    \draw[flow, draw=cGreen!60!black, line width=0.7pt] ($(P3.north west)+(1.75,-2.48)$)--($(finaldiff.west)+(0,0.30)$);
    \node[chip=cGreen, anchor=north, font=\sffamily\bfseries\fontsize{6.8}{7.6}\selectfont]
       (auditc) at ($(finaldiff.south)+(0,-0.16)$) {\faCheckDouble\,audit pass};
    \draw[flow, draw=cGreen!60!black, line width=0.7pt] (finaldiff.south)--(auditc.north);

    \foreach \i/\tcol/\tlb in {0/cTeal/{link}, 1/cAmber/{$N^\star$}, 2/cGreen/{atoms}}{
      \coordinate (gA) at ($(P\i.east)+(0.04,0)$ |- {($(P0.north)+(0,\midy)$)});
      \coordinate (gB) at ($(P\the\numexpr\i+1\relax.west)+(-0.04,0)$ |- {($(P0.north)+(0,\midy)$)});
      \draw[flow] (gA)--(gB);
      \node[font=\sffamily\fontsize{6.4}{7}\selectfont, text=\tcol!72!black, anchor=south] at ($(gA)!0.5!(gB)+(0,0.07)$) {\tlb};
    }

  \end{tikzpicture}%
  }
  \caption{\method{} turns a fixed candidate pool into one auditable final patch in four deterministic steps.}
  \label{fig:method-overview}
\end{figure*}

\method turns a fixed candidate pool into one auditable final patch in four deterministic steps (Figure~\ref{fig:method-overview}): (1)~\emph{candidate view and exact support} collapses exact-duplicate diffs into support counts; (2)~\emph{repair-neighborhood fusion} groups candidates that touch related files and share edit vocabulary; (3)~\emph{representative selection} chooses one representative under the pool structure, admitting a local edit only under a same-surface guard; and (4)~\emph{evidence-constrained fusion} (ECF) keeps the edit atoms repeated across candidates and prunes unsupported ones.

\subsection{Step 1: Candidate View and Exact Support}
\label{sec:candidate-interface}

Step~1 reduces the raw pool to what the selector may inspect at decision time, and records how often each distinct fix recurs.
The input is an unordered candidate pool for a bug, and every evaluated pool exposes unified diffs.
From these \method builds a decision-time view of diff-derived fields, duplicate counts, and unlabeled pool structure only, excluding validation outcomes and run-quality metadata.
It then collapses exact-duplicate diffs into a distinct-patch pool $\widetilde{\mathcal{P}}_b$, giving each distinct candidate a support count equal to the number of raw candidates that share its canonical diff; content-different diffs stay distinct even when they are semantically similar.

\subsection{Step 2: Repair-Neighborhood Fusion}

Step~2 groups candidates that agree on \emph{what} to change into repair neighborhoods, then keeps the grouping the pool most agrees on.
Agreement between two candidates is measured on the files they touch and the identifier-like tokens they edit.
For a candidate $p$, let $F(p)$ be its touched-file set and $T(p)$ the identifier-like token set from its unified diff; for a pair $p_i,p_j$, file and token overlap are the Jaccard similarities
\begin{equation}
J_F(i,j)=\frac{|F(p_i)\cap F(p_j)|}{|F(p_i)\cup F(p_j)|},\quad
J_T(i,j)=\frac{|T(p_i)\cap T(p_j)|}{|T(p_i)\cup T(p_j)|}.
\end{equation}
Pairs that share no file evidence or no token evidence are never linked.
For the rest, \method avoids a fixed similarity cutoff: it considers only the overlap graphs induced by the distinct overlap levels actually present in this pool, a finite set $\mathcal{G}_b$ of candidate granularities for bug $b$.
Among these it keeps the granularity whose induced neighborhoods make the Step-3 routing decision most decisive (made precise in Step~3), and takes the connected components of that graph as the repair neighborhoods.

\method then scores each neighborhood. For a neighborhood $N$ it computes four static signals,
\begin{equation}
\begin{aligned}
\mathrm{Sup}(N)&=\sum_{p\in N}\mathrm{sup}(p),\\
\mathrm{Var}(N)&=|N|,\\
\mathrm{Coh}(N)&=\frac{1}{|N|(|N|-1)}
  \sum_{\substack{p,q\in N\\p\ne q}}J_T(p,q),\\
\mathrm{Conc}(N)&=-|\{F(p)\mid p\in N\}|,
\end{aligned}
\end{equation}
with $\mathrm{Coh}(N)=0$ for singletons, capturing how much pooled support, how many variants, how cohesive, and how file-concentrated the neighborhood is.
Neighborhoods are ranked by a pairwise majority-Copeland vote over the four signals: the neighborhood that is larger on more of them wins each pairwise comparison, scored as wins minus losses, with ties broken by the ordinal signal tuple and a canonical-diff hash.
\method takes the top-ranked neighborhood $N^\star$, the repair surface on which all later representative changes must stay.

\subsection{Step 3: Representative Selection}
\label{sec:representative-selection}

Step~3 picks one auditable representative inside the top-ranked neighborhood $N^\star$, optionally upgrades it to a stronger local edit, and routes between these two options by pool structure alone.
The default representative balances exact support with token centrality; for a candidate $p$,
\begin{equation}
C_T(p)=\frac{1}{|\widetilde{\mathcal{P}}_b|-1}\sum_{q\in\widetilde{\mathcal{P}}_b,q\ne p}J_T(p,q),
\end{equation}
and \method ranks members of $N^\star$ by a pairwise majority vote over $\mathrm{sup}(p)$ and $C_T(p)$.
It also considers replacing this representative with a stronger \emph{local edit candidate} from the same neighborhood, admitted only under the \emph{same-surface guard}: a local candidate $p$ is eligible only when it stays on the representative's surface, with its touched-file set contained in that of the current representative $c$ ($F(p)\subseteq F(c)$, an equal surface or a strict contraction) or confined to the same subsystem, so it never moves to an unrelated file.
Among eligible local candidates, \method takes the best under a lexicographic support order (file-surface support, edit-region support, edit-token centrality, exact support, compactness, deterministic hash) and replaces the representative only when that candidate is strictly better.

The choice between the neighborhood representative and the local edit is routed by two pool-level confidences.
Let $\mu$ be the mean token centrality over the distinct pool and $\phi$ the repair-neighborhood fragmentation ratio, the number of induced neighborhoods divided by the number of distinct candidates; then
\begin{equation}
  C_{\mathrm{nbr}}=\mu(1-\phi),\qquad
  C_{\mathrm{loc}}=(1-\mu)\phi .
\end{equation}
$C_{\mathrm{nbr}}$ is high when a token-central pool fragments into few neighborhoods, and $C_{\mathrm{loc}}$ when a token-diffuse pool fragments into many.
The fragmentation $\phi$, and hence both confidences, depends on the overlap graph, so we write $C_{\mathrm{nbr}}(G)$ and $C_{\mathrm{loc}}(G)$ when the graph matters.
\method takes the neighborhood-first representative when $C_{\mathrm{nbr}}\ge C_{\mathrm{loc}}$ and the same-surface local edit candidate otherwise.
The same gap also fixes the Step-2 granularity: among the candidate graphs $\mathcal{G}_b$, \method keeps the view $G^\star=\arg\max_{G\in\mathcal{G}_b}|C_{\mathrm{nbr}}(G)-C_{\mathrm{loc}}(G)|$, the one whose routing is most decisive.
In this diff-only path the chosen output is still one submitted candidate diff from the pool.

\subsection{Step 4: Evidence-Constrained Fusion}

Step~4 is the content-level stage: it rebuilds the representative from the edit atoms that recur across candidates, drops the unsupported ones, and accepts the result only when it clears the static acceptance guards.
\method decomposes each candidate into scope-local edit atoms relative to the buggy scope, where an atom $a=(I,R)$ pairs a contiguous buggy-scope interval $I$ with its replacement text $R$.
Atoms vote only within the same scope, so edits from different methods, classes, or modules never replace one another.
These scopes are built with language-aware parsers from candidate method bodies or AST boundaries in the checked-out repository, a Python AST locator and tree-sitter grammars for the other languages, falling back to a brace-balanced block only when a scope cannot be parsed.

Let $\mathcal{S}_b$ be the repair scopes for bug $b$, and for a candidate $p$ and scope $s\in\mathcal{S}_b$ let $A_s(p)$ be its atom set; an atom's candidate-pool support is
\begin{equation}
  \mathrm{sup}_s(a)=|\{p\in\widetilde{\mathcal{P}}_b \mid a\in A_s(p)\}|,
\end{equation}
and $\mathcal{A}_s=\bigcup_{p\in\widetilde{\mathcal{P}}_b}A_s(p)$ is the scope's atom vocabulary.
For the selected representative $c$, ECF admits an atom from $\mathcal{A}_s$ only when it is \emph{scope-compatible} with $c$, meaning it edits a scope already touched by $c$ and does not overlap an accepted atom on the same buggy interval, and is supported by another candidate.
The fused atom set is chosen by a lexicographic evidence order rather than a weighted score: ECF first requires compatible, cross-candidate-supported atoms, then prefers larger aggregate support, and uses compactness only to break support-equivalent choices.

ECF accepts a reconstructed patch in one of two modes, both keeping only cross-candidate-supported atoms.
The default \emph{contraction} mode rewrites a single scope and is accepted only when it is non-expanding, strictly shorter, and evidence-dominant:
\begin{equation}
  \begin{aligned}
  &\mathrm{Surf}(\mathrm{ECF}(c))\subseteq \mathrm{Surf}(c),\\
  &\mathrm{tok}(\mathrm{ECF}(c)) < \mathrm{tok}(c),\\
  &\mathrm{ASup}(\mathrm{ECF}(c))>\mathrm{ASup}(c)
    \quad\text{or}\quad
    \mathrm{Exact}(\mathrm{ECF}(c))>1 ,
  \end{aligned}
\end{equation}
where $\mathrm{Surf}$ is the touched scope/file surface, $\mathrm{ASup}$ sums edit-atom support, and $\mathrm{Exact}$ counts exact candidate support when the reconstruction maps to an existing candidate.
When the supported repair instead spans several scopes, ECF forms a \emph{supported scope-union}: it unions up to $K=3$ non-overlapping scope bodies whose edit atoms are each repeated across at least 2 candidates, accepted only under a compactness gate (bounded fused tokens and an edit-token overlap with the base revision below $0.8$).

\section{Experimental Setup}\label{sec:experimental-setup}

\subsection{Task and Decision-Time Boundary}

We study fixed candidate-pool patch fusion after multiple repair runs have already produced candidate patches for the same bug.
For a bug $b$, let $\mathcal{R}=\{r_1,\dots,r_m\}$ be the fixed set of available repair runs and let $\mathcal{P}^{\mathrm{raw}}_b=\{p_{b,r}\mid r\in\mathcal{R}\}$ be the raw candidate pool.
Because different runs can submit the same unified diff, \method collapses exact clones into a distinct-patch candidate pool $\widetilde{\mathcal{P}}_b$ before selecting or reconstructing one final patch $\tilde{p}^{\star}_b$.
At decision time the selector uses no test outcome or run metadata: it may inspect candidate artifacts (unified diffs, hashes, and exposed method bodies), exact duplicate counts, and unlabeled pool structure (file and token overlap and the induced repair neighborhoods), but never test outcomes, official solved labels, leaderboard rank or run pass rate, or run identity and source order.
Outcome labels are attached only after the final patch is produced.

\subsection{Benchmark}

We construct \dataset, a reusable patch-fusion benchmark that assembles fixed multi-system candidate pools from public leaderboard and repair runs on SWE-bench Verified, SWE-bench Multilingual, and Defects4J, each pool paired with cached official outcome labels that score any pooled candidate without rerunning a test, while patches generated outside the pool are scored by the official oracle.
Tables use benchmark-facing names for its three pools: \textbf{SWE-bench Verified} is the 500-bug official Verified pool, \textbf{SWE-bench Multilingual} the 300-bug public multilingual pool, and \textbf{Defects4J} the primary Defects4J pass@10 pool.

We evaluate on two fixed cross-run SWE-bench candidate pools.
SWE-bench Verified~\cite{jimenez2024swebench} uses the submitted patch predictions from 6 official SWE-bench Verified entries on the public leaderboard~\cite{swebenchLeaderboard2026}: Live-SWE-agent with Claude Opus 4.5 and with Gemini 3 Pro Preview~\cite{xia2025livesweagent}, TRAE with Doubao-Seed-Code~\cite{bytedance2025traeagent}, Atlassian Rovo Dev~\cite{atlassian2025rovodev}, EPAM AI/Run Developer Agent with Claude 4 Sonnet~\cite{epam2025airun}, and ACoder~\cite{acoder2025swebench}.
SWE-bench Multilingual~\cite{swebenchLeaderboard2026} is formed from 7 public repair runs: Gemini 3 Flash, Claude Opus 4.6, Claude Opus 4.5, GLM-5, Gemini 3 Pro, MiniMax 2.5, and Kimi K2.5.
These are the first-page leaderboard entries (snapshot April 2026) that provide downloadable, format-compatible patch predictions; first-page entries without usable prediction files are excluded.
We additionally evaluate on Defects4J \cite{just2014defects4j} by reusing Defects4J pass@10 candidate pools produced by MORepair~\cite{yang2024morepair}, which contain 371 Java bugs and 10 generated candidate repairs per bug.

We release each source's candidate count, duplicate rate, and per-source statistics with the artifact; where a source has an empty or missing prediction for an instance, that single entry is dropped only for that source.

\subsection{Baselines}\label{sec:exp-baselines}

Our main comparisons use \emph{matched candidate-pool selectors}: each receives the same unordered candidate pool as \method and obeys the same decision-time boundary, so any difference reflects the selection rule rather than the generation stack.
Table~\ref{tab:baseline-taxonomy} groups the baselines into families of post-generation selectors by granularity and decision-time properties; whole-diff selectors dominate, and among the families we evaluate, only \method fuses edit atoms from within candidate patches while staying test-free, model-free, and deterministic.
\begin{table}[t]
  \centering
  \caption{Families of post-generation patch selectors and their decision-time properties.}
  \label{tab:baseline-taxonomy}
  \footnotesize
  \setlength{\tabcolsep}{4pt}
  \renewcommand{\arraystretch}{1.12}
  \resizebox{\columnwidth}{!}{%
  \begin{tabular}{@{}llcccl@{}}
    \toprule
    Family & Granularity & Test-free & Model-free & Determ. & Representative \\
    \midrule
    Minimization & whole patch & \checkmark & \checkmark & \checkmark & Minimal-change~\cite{legoues2012genprog} \\
    Consensus & whole patch & \checkmark & \checkmark & \checkmark & Token medoid~\cite{li2022alphacode} \\
    Rank aggregation & whole patch & \checkmark & \checkmark & \checkmark & Borda/Copeland/MC4~\cite{dwork2001rankaggregation} \\
    Learned ranker & whole patch & \checkmark & $\times$ & \checkmark & LM naturalness~\cite{kang2022lmpatch} \\
    LLM-as-judge & whole patch & \checkmark & $\times$ & $\times$ & DeepSeek-V4-Pro listwise \\
    LLM-generated & free-form & \checkmark & $\times$ & $\times$ & DeepSeek-V4-Pro fusion \\
    Test-based & whole patch & $\times$ & \checkmark & \checkmark & Agentless~\cite{xia2025agentless} \\
    \midrule
    \textbf{Evidence fusion} & \textbf{edit atoms} & \checkmark & \checkmark & \checkmark & \textbf{\method{} (ours)} \\
    \bottomrule
  \end{tabular}}
\end{table}

Two accounting rows frame the pool, the uniform-random expected solved count and the best single source, the latter a post-hoc reference because identifying it needs labels unavailable at decision time.
The deterministic selectors are minimal-change (the shortest, new-file-avoiding patch, an APR minimization control~\cite{legoues2012genprog}), token medoid (the distinct patch most central to the pool by token overlap), hunk consensus (the candidate with the strongest changed-line support, without \method{}'s neighborhoods or routing), the Borda, Copeland, and MC4 rank-aggregation controls~\cite{dwork2001rankaggregation}, in which each candidate ranks the others by patch similarity weighted by its run support and the three methods aggregate those rankings (MC4 with a damped Markov chain), and the Agentless exact-identity vote~\cite{xia2025agentless}.
The LM naturalness ranker is a frozen CodeGen-350M model that picks the candidate with the lowest mean negative log-likelihood over added lines, implementing LM-based patch prioritization~\cite{kang2022lmpatch} under the naturalness hypothesis for software~\cite{hindle2012naturalness}.
We deliberately exclude supervised learned patch-correctness predictors~\cite{xiong2018patchcorrectness,ye2021patchassessment,tang2024patchrepr} from the matched selectors, because training one needs labeled correct and incorrect patches that for these benchmarks would have to come either from the benchmarks themselves (contamination) or from an off-distribution released checkpoint, whereas the naturalness ranker is the learned-model selector that needs no such labels.
Two model-based selectors run a DeepSeek-V4-Pro endpoint (temperature 0, candidates in hash order, JSON output): a listwise judge that chooses one submitted candidate, and a free-form fusion baseline that instead asks the model to emit a final patch.

As a single \emph{test-based} reference point, the \emph{Agentless} row restricts the exact-identity vote to candidates that pass the benchmark's regression tests (the \texttt{PASS\_TO\_PASS} tests on SWE-bench and the relevant tests in the reused Defects4J validation), voting over the survivors; its test-free counterpart is the \emph{Agentless w/o test} row.
It is the only row that consults test outcomes (one regression run per candidate), so it sits outside the matched test-free selector family, and we report it only to check whether test access lets a strong baseline beat the test-free \method.

\subsection{Evaluation Metrics and Protocol}

Every selector emits one final patch per bug.
We evaluate that final patch with each benchmark's official outcome oracle, never a self-reported one: exact submitted candidates carry the official leaderboard evaluation of their run, and the new patches ECF emits are scored by the same official SWE-bench runner.
On SWE-bench, the primary metric is \emph{solved count}: the number of benchmark instances whose final patch passes the official SWE-bench evaluation.
On Defects4J, the primary metric is \emph{plausible count}: the number of final patches that pass the stored Defects4J validation tests in the reused MORepair artifacts.
For a benchmark $B$ with $N$ bugs, let $p_b^S$ be the patch emitted by selector $S$ for bug $b$, and let $y_b(p)\in\{0,1\}$ denote the post-hoc benchmark outcome label.
The headline count is:
\begin{equation}
\mathrm{Count}(S)=\sum_{b\in B} y_b(p_b^S).
\end{equation}
The uniform-random expectation is computed on the same distinct candidate pool:
\begin{equation}
\mathbb{E}[\mathrm{Rand}]=
\sum_{b\in B}\frac{1}{|\widetilde{\mathcal{P}}_b|}
\sum_{p\in\widetilde{\mathcal{P}}_b} y_b(p).
\end{equation}
In the main fixed candidate-pool comparison, $\Delta_R$ is the count difference from this expectation:
\begin{equation}
\Delta_R(S)=\mathrm{Count}(S)-\mathbb{E}[\mathrm{Rand}].
\end{equation}

For paired comparisons, a win is an instance solved by \method but not by the comparator, and a loss is an instance solved by the comparator but not by \method.
Instances where both methods have the same outcome are paired ties and are excluded from the W/L test statistic.
For a comparator $C$, the paired counts are:
\begin{align}
W(S,C)&=|\{b\mid y_b(p_b^S)=1\land y_b(p_b^C)=0\}|,\\
L(S,C)&=|\{b\mid y_b(p_b^S)=0\land y_b(p_b^C)=1\}|.
\end{align}
We report W/L counts and use an exact two-sided paired signed test over these outcome-discordant instances.
Holm-Bonferroni correction is applied within each planned matched candidate-pool paired-baseline family for each SWE-bench candidate pool, with the corrected per-pool tests released in the artifact.
Uniform-random and best-single-source are accounting quantities, not deployable selectors, so they are excluded from the matched candidate-pool paired family; we additionally report a per-instance comparison of \method against the best single source as a complementarity check (Section~\ref{sec:results}).

\subsection{Research Questions}

\textbf{RQ-1: How effective is \method for final-patch production from fixed candidate pools?}
We compare \method with matched candidate-pool selectors on \dataset under the same decision-time boundary, reporting whole-benchmark counts and paired W/L tests against the ranking, consensus, Agentless, LM-naturalness, and DeepSeek-V4-Pro baselines, and analyzing why the same model's free-form fusion underperforms its selection.

\textbf{RQ-2: Which components account for the gains, and where do residual misses remain?}
RQ-2.1 ablates each design choice (repair-neighborhood fusion, local edit evidence, the same-surface guard, pool-level routing, and ECF) one at a time on the same pools.
RQ-2.2 diagnoses the candidate-reachable bugs that the routed pre-ECF representative still misses.

\textbf{RQ-3: What does the post-generation decision cost, and how does it compare with other selectors?}
We place every baseline on a cost-versus-repairs plane, comparing the deterministic selectors' post-generation wall-clock per bug against the latency and dollar cost of the model-based rows.

\textbf{RQ-4: How robust is \method to pool composition?}
We subsample each pool and track the fraction of candidate-reachable bugs solved at each pool size, separating system diversity from sampling randomness.

\section{Experiments \& Results}\label{sec:results}

\begin{table*}[t]
  \centering
  \caption{Solved counts, decision-time latency, and pooled paired W/L with $p$ versus \method on \dataset.}
  \label{tab:main-fixed-pool}
  \tiny
  \begingroup
  \renewcommand{\arraystretch}{1.10}
  \setlength{\tabcolsep}{1.5pt}
  \resizebox{\textwidth}{!}{%
  \begin{tabular}{lcccccccccccccc}
    \toprule
    \multirow{2}{*}{\shortstack{\\\\Approach}} &
    \multirow{2}{*}{\shortstack{\\\\Time\\ms / bug}} &
    \multirow{2}{*}{\shortstack{\\SWE-bench\\Verified\\(500 bugs)}} &
    \multicolumn{8}{c}{\shortstack{SWE-bench Multilingual (300 bugs)}} &
    \multirow{2}{*}{\shortstack{\\\\Defects4J\\(371 bugs)}} &
    \multicolumn{2}{c}{\shortstack{vs \method}} \\
    \cmidrule(lr){4-11}\cmidrule(lr){13-14}
    & & & \shortstack{All\\300} & \shortstack{Java\\43} & \shortstack{Rust\\43} & \shortstack{JS/TS\\43} & \shortstack{PHP\\43} & \shortstack{Go\\42} & \shortstack{Ruby\\44} & \shortstack{C/C++\\42} & & W/L & $p$ \\
    \midrule
    Random Expectation & -- & 388.0 & 212.3 & 34.0 & 33.3 & 30.3 & 30.5 & 25.1 & 28.6 & 30.6 & 47.3 & -- & -- \\
    Best single source & -- & \scoregain{396}{\gainpos{+8.0}} & \scoregain{218}{\gainpos{+5.7}} & \scoregain{35}{\gainpos{+1.0}} & \scoregain{35}{\gainpos{+1.7}} & \scoregain{28}{\gainneg{-2.3}} & \scoregain{30}{\gainneg{-0.5}} & \scoregain{26}{\gainpos{+0.9}} & \scoregain{30}{\gainpos{+1.4}} & \scoregain{34}{\gainpos{+3.4}} & \scoregain{62}{\gainpos{+14.7}} & -- & -- \\
    \midrule
    Minimal-change~\cite{legoues2012genprog} & 0.02 & \scoregain{382}{\gainneg{-6.0}} & \scoregain{209}{\gainneg{-3.3}} & \scoregain{33}{\gainneg{-1.0}} & \scoregain{31}{\gainneg{-2.3}} & \scoregain{32}{\gainpos{+1.7}} & \scoregain{30}{\gainneg{-0.5}} & \scoregain{24}{\gainneg{-1.1}} & \scoregain{28}{\gainneg{-0.6}} & \scoregain{31}{\gainpos{+0.4}} & \scoregain{62}{\gainpos{+14.7}} & 120/24 & $<0.05$ \\
    Token medoid~\cite{li2022alphacode} & 0.02 & \scoregain{403}{\gainpos{+15.0}} & \scoregain{219}{\gainpos{+6.7}} & \scoregain{37}{\gainpos{+3.0}} & \scoregain{35}{\gainpos{+1.7}} & \scoregain{30}{\gainneg{-0.3}} & \scoregain{32}{\gainpos{+1.5}} & \scoregain{25}{\gainneg{-0.1}} & \scoregain{30}{\gainpos{+1.4}} & \scoregain{30}{\gainneg{-0.6}} & \scoregain{73}{\gainpos{+25.7}} & 60/6 & $<0.05$ \\
    Hunk consensus & 0.03 & \scoregain{397}{\gainpos{+9.0}} & \scoregain{217}{\gainpos{+4.7}} & \scoregain{35}{\gainpos{+1.0}} & \scoregain{36}{\gainpos{+2.7}} & \scoregain{32}{\gainpos{+1.7}} & \scoregain{31}{\gainpos{+0.5}} & \scoregain{27}{\gainpos{+1.9}} & \scoregain{27}{\gainneg{-1.6}} & \scoregain{29}{\gainneg{-1.6}} & \scoregain{65}{\gainpos{+17.7}} & 86/16 & $<0.05$ \\
    Patch Borda~\cite{dwork2001rankaggregation} & 0.22 & \scoregain{403}{\gainpos{+15.0}} & \scoregain{219}{\gainpos{+6.7}} & \scoregain{37}{\gainpos{+3.0}} & \scoregain{35}{\gainpos{+1.7}} & \scoregain{30}{\gainneg{-0.3}} & \scoregain{32}{\gainpos{+1.5}} & \scoregain{25}{\gainneg{-0.1}} & \scoregain{30}{\gainpos{+1.4}} & \scoregain{30}{\gainneg{-0.6}} & \scoregain{70}{\gainpos{+22.7}} & 63/6 & $<0.05$ \\
    Patch Copeland~\cite{dwork2001rankaggregation} & 0.25 & \scoregain{403}{\gainpos{+15.0}} & \scoregain{218}{\gainpos{+5.7}} & \scoregain{36}{\gainpos{+2.0}} & \scoregain{34}{\gainpos{+0.7}} & \scoregain{30}{\gainneg{-0.3}} & \scoregain{32}{\gainpos{+1.5}} & \scoregain{26}{\gainpos{+0.9}} & \scoregain{30}{\gainpos{+1.4}} & \scoregain{30}{\gainneg{-0.6}} & \scoregain{71}{\gainpos{+23.7}} & 65/8 & $<0.05$ \\
    Patch MC4~\cite{dwork2001rankaggregation} & 0.53 & \scoregain{400}{\gainpos{+12.0}} & \scoregain{217}{\gainpos{+4.7}} & \scoregain{36}{\gainpos{+2.0}} & \scoregain{33}{\gainneg{-0.3}} & \scoregain{30}{\gainneg{-0.3}} & \scoregain{32}{\gainpos{+1.5}} & \scoregain{26}{\gainpos{+0.9}} & \scoregain{30}{\gainpos{+1.4}} & \scoregain{30}{\gainneg{-0.6}} & \scoregain{70}{\gainpos{+22.7}} & 69/7 & $<0.05$ \\
    LM naturalness ranker~\cite{kang2022lmpatch} & 77.1 & \scoregain{390}{\gainpos{+2.0}} & \scoregain{211}{\gainneg{-1.3}} & \scoregain{33}{\gainneg{-1.0}} & \scoregain{34}{\gainpos{+0.7}} & \scoregain{30}{\gainneg{-0.3}} & \scoregain{30}{\gainneg{-0.5}} & \scoregain{23}{\gainneg{-2.1}} & \scoregain{30}{\gainpos{+1.4}} & \scoregain{31}{\gainpos{+0.4}} & \scoregain{30}{\gainneg{-17.3}} & 141/23 & $<0.05$ \\
    DeepSeek-V4-Pro listwise & 1260 & \scoregain{396}{\gainpos{+8.0}} & \scoregain{214}{\gainpos{+1.7}} & \scoregain{35}{\gainpos{+1.0}} & \scoregain{34}{\gainpos{+0.7}} & \scoregain{28}{\gainneg{-2.3}} & \scoregain{35}{\gainpos{+4.5}} & \scoregain{26}{\gainpos{+0.9}} & \scoregain{28}{\gainneg{-0.6}} & \scoregain{28}{\gainneg{-2.6}} & \scoregain{74}{\gainpos{+26.7}} & 90/25 & $<0.05$ \\
    DeepSeek-V4-Pro fusion & 6024 & \scoregain{317}{\gainneg{-71.0}} & \scoregain{183}{\gainneg{-29.3}} & \scoregain{35}{\gainpos{+1.0}} & \scoregain{26}{\gainneg{-7.3}} & \scoregain{27}{\gainneg{-3.3}} & \scoregain{27}{\gainneg{-3.5}} & \scoregain{19}{\gainneg{-6.1}} & \scoregain{23}{\gainneg{-5.6}} & \scoregain{26}{\gainneg{-4.6}} & \scoregain{40}{\gainneg{-7.3}} & 221/16 & $<0.05$ \\
    \midrule
    Agentless & 386108 & \scoregain{392}{\gainpos{+4.0}} & \scoregain{222}{\gainpos{+9.7}} & \scoregain{38}{\gainpos{+4.0}} & \scoregain{33}{\gainneg{-0.3}} & \scoregain{32}{\gainpos{+1.7}} & \scoregain{31}{\gainpos{+0.5}} & \scoregain{27}{\gainpos{+1.9}} & \scoregain{31}{\gainpos{+2.4}} & \scoregain{30}{\gainneg{-0.6}} & \scoregain{84}{\gainpos{+36.7}} & 80/33 & $<0.05$ \\
    Agentless w/o test~\cite{xia2025agentless} & 0.07 & \scoregain{384}{\gainneg{-4.0}} & \scoregain{217}{\gainpos{+4.7}} & \scoregain{37}{\gainpos{+3.0}} & \scoregain{32}{\gainneg{-1.3}} & \scoregain{32}{\gainpos{+1.7}} & \scoregain{30}{\gainneg{-0.5}} & \scoregain{26}{\gainpos{+0.9}} & \scoregain{30}{\gainpos{+1.4}} & \scoregain{30}{\gainneg{-0.6}} & \scoregain{74}{\gainpos{+26.7}} & 96/22 & $<0.05$ \\
    \midrule
    \textbf{\method{}} & \textbf{3.28} & \textbf{\scoregain{426}{\gainpos{+38.0}}} & \textbf{\scoregain{236}{\gainpos{+23.7}}} & \textbf{\scoregain{40}{\gainpos{+6.0}}} & \textbf{\scoregain{38}{\gainpos{+4.7}}} & \textbf{\scoregain{33}{\gainpos{+2.7}}} & \textbf{\scoregain{33}{\gainpos{+2.5}}} & \textbf{\scoregain{27}{\gainpos{+1.9}}} & \textbf{\scoregain{33}{\gainpos{+4.4}}} & \textbf{\scoregain{32}{\gainpos{+1.4}}} & \textbf{\scoregain{87}{\gainpos{+39.7}}} & -- & -- \\
    \bottomrule
  \end{tabular}  }
  \endgroup
\end{table*}

\subsection{RQ-1: Overall Fixed-Pool Results}

\noindent\textbf{[Experimental Design]}:
We run the full \method and the matched candidate-pool selectors on the three complete \dataset pools under the decision-time boundary of Section~\ref{sec:experimental-setup}, reporting each whole-benchmark solved count with its gain $\Delta_R$ over uniform-random selection (Table~\ref{tab:main-fixed-pool}).
We compare \method against every baseline with W/L counts and exact paired sign tests, noting that the DeepSeek-V4-Pro fusion row generates a final patch rather than selecting one from the pool.

\noindent\textbf{[Experimental Results]}:
Three patterns are visible in Table~\ref{tab:main-fixed-pool}.
First, \method solves the most bugs among matched candidate-pool selectors across all three benchmark settings, adding 38.0, 23.7, and 39.7 bugs over uniform random selection on SWE-bench Verified, SWE-bench Multilingual, and Defects4J.
Its strongest competitor is token medoid, which \method beats by 23, 17, and 14 bugs, with a combined 60/6 discordant-pair record across the three pools.
The margins over the remaining selectors (rank aggregation, hunk consensus, Agentless w/o test, the LM naturalness ranker, and DeepSeek-V4-Pro listwise) are read directly from Table~\ref{tab:main-fixed-pool}.
Exact two-sided paired sign tests over the outcome-discordant instances pooled across all three benchmarks (the W/L and $p$ columns of Table~\ref{tab:main-fixed-pool}) show statistically significant gaps: \method beats all 9 matched test-free selectors at $p<0.05$ on the pooled sign test, and each per-pool gap also holds under Holm-Bonferroni correction (released with the artifact).
The per-pool W/L counts and confidence intervals are released with the artifact.

The best single-source row is a post-hoc reference, not a deployable selector.
\method is also above this reference on all three benchmark settings, but the main deployable comparison is the gap over random selection and matched candidate-pool selectors.
Per instance, \method solves 41 Verified and 27 Multilingual bugs that the best single source leaves unsolved while losing only 11 and 9 (net $+30$ and $+18$, $p<0.05$ paired sign test), recovering repairs no single system supplies.
For reference, a post-hoc oracle that keeps, for each bug, a candidate carrying the official solved label reaches the candidate-reachable ceiling ($443$ and $263$ on the two SWE-bench pools); \method approaches this ceiling ($426$ and $236$) with no label or test consulted at decision time.

We further probe whether test access alone would let a strong baseline catch \method.
The \emph{Agentless} row in Table~\ref{tab:main-fixed-pool} keeps only candidates that pass the regression tests before the exact-identity vote, lifting the test-free \emph{Agentless w/o test} from $384/217/74$ to $392/222/84$ on the three pools.
Even with this test signal it trails \method by $34$, $14$, and $3$ bugs and loses the pooled paired test ($80/33$, $p<0.05$); the lift is largest on Defects4J, where roughly half of the candidates fail to compile and regression filtering removes the most failing candidates, yet test-free deterministic fusion still leads.
Regression access thus narrows but never closes the gap, while forfeiting the decision-time, test-free property that keeps \method deployable: the row must run one regression suite per candidate, an estimated $386{,}108$\,ms per bug.

Second, DeepSeek-V4-Pro rows provide model-based post-generation baselines under the same decision-time boundary.
The listwise judge remains below \method on all three benchmark settings while requiring model inference rather than deterministic candidate-pool evidence fusion.
The free-form fusion row asks the model to generate a final patch rather than select one from the pool; even so, \method beats it $221/16$ over the pooled discordant instances ($p<0.05$).
Here, for SWE-bench we use official-runner outcomes for new diffs; for Defects4J, we keep the row candidate-aligned and only reuse cached labels for generated methods that match existing candidates.
This is a different operating point from \method: it spends model inference time generating patches, whereas \method fuses static candidate evidence into an auditable first patch.
Their decision-stage cost, where \method holds a further advantage, is quantified in RQ-3.

Third, the Multilingual language slices serve as a robustness check rather than as a basis for per-language claims.
Each slice contains only 42-44 bugs, so we do not consider individual one- or two-bug differences significant.
Across the slices, though, one pattern is consistent: \method is the only method whose per-language $\Delta_R$ stays positive on every language, whereas every baseline regresses below uniform random on at least one, including the strongest baseline, token medoid, on JS/TS, Go, and C/C++.
This uniform sign indicates that \method's gain is broad across languages rather than concentrated in a few.

\subsubsection{Selection versus generation: why free-form fusion loses}
\begin{table}[h]
  \centering
  \caption{DeepSeek-V4-Pro as a selector versus a free-form generator.}
  \label{tab:fusion-failure}
  \setlength{\tabcolsep}{4pt}
  \renewcommand{\arraystretch}{1.12}
  \begin{tabular}{lcc}
    \toprule
    DeepSeek-V4-Pro operating mode & Verified & Multilingual \\
    \midrule
    Listwise (selects one candidate) & 396/500 & 214/300 \\
    Free-form fusion (generates a patch) & 317/500 & 183/300 \\
    \midrule
    \multicolumn{3}{c}{\emph{Where fusion loses}} \\
    Invalid, non-applying generated diffs & 96 & 38 \\
    Just reproduces an existing candidate & 87 & 144 \\
    Reachable bugs left unsolved & 126/443 & 80/263 \\
    \quad invented a failing novel patch & 92 & 49 \\
    \bottomrule
  \end{tabular}
\end{table}

The DeepSeek-V4-Pro rows let us compare selection and generation with one model held fixed: the listwise judge selects a submitted candidate, while the fusion baseline asks the same model to emit a final patch.
Selection wins by 79 and 31 bugs (Table~\ref{tab:fusion-failure}): the same model solves 396 and 214 by choosing versus 317 and 183 by generating.
Fusion loses because generation re-opens failure modes selection eliminates by construction.
First, 96 of 500 Verified diffs (and 38 of 300 Multilingual) do not even apply: the model emits malformed or mis-anchored hunks, often wrapped in a markdown or JSON envelope, that a selector can never produce.
Second, on bugs a candidate already solves, the model still leaves 126/443 (Verified) and 80/263 (Multilingual) unsolved, and the main cause is self-inflicted: in 92 and 49 of these it generates a new patch that fails rather than choosing the correct candidate already in the pool.
\texttt{apache\_\_druid-14136} is typical: the model wraps its diff in a JSON envelope and proposes a plausible empty-interval guard that the tests reject, while a solving candidate was available.
Half the time on Multilingual ($144/300$) it simply reproduces an existing candidate, so even many of its successes are mere selection.
Once correct patches already populate the pool, handing the decision to a strong generator is therefore worse than a deterministic selector: generation adds format errors, hallucinated edits, and the risk of overwriting an available fix.

\findx{\findtag{[RQ-1] Findings.} (1) \method comes within 17 of the candidate-reachable ceiling on Verified (426/500) and beats every matched selector at $p<0.05$.
(2) Per instance it recovers repairs no single source solves (net $+30$ and $+18$ bugs).
(3) Given the same decision, the evaluated DeepSeek-V4-Pro baseline does worse, not better: its free-form fusion loses 79 and 31 bugs, mostly by overwriting a correct candidate already in the pool.
\findtag{Insight.} When a fixed pool already holds correct patches, the bottleneck moves from generation to \emph{selection}; in this setting, a stronger generator does not help and can hurt.}

\subsection{RQ-2: Component Contributions and Residual Misses}

\noindent\textbf{[Experimental Design]}:
We ablate \method one internal design choice at a time on the same candidate pools, keeping candidate canonicalization and exact-support counting fixed.
The `w/o ECF' row disables the whole content-fusion module, so it measures ECF's marginal contribution inside the same pipeline; the other rows each alter one pre-ECF decision that supplies ECF with a representative.

\begin{table}[h]
  \centering
  \caption{Ablation of \method components.}
  \label{tab:component-ablation}
  \scriptsize
  \setlength{\tabcolsep}{3pt}
  \renewcommand{\arraystretch}{1.08}
  \begin{tabular}{@{}p{52mm}ccc@{}}
    \toprule
    Variant & Verified & Multi. & Defects4J \\
    \midrule
    \method & \textbf{426} & \textbf{236} & \textbf{87} \\
    \midrule
    Step 2: \method{} w/o repair-neighborhood & 421 & 214 & 62 \\
    \midrule
    Step 3: \method{} w/o local edit evidence & 405 & 230 & 78 \\
    Step 3: \method{} w/o same-surface guard & 411 & 214 & 65 \\
    Step 3: \method{} w/o pool-level route & 421 & 230 & 78 \\
    \midrule
    Step 4: \method{} w/o ECF & 421 & 230 & 78 \\
    \bottomrule
  \end{tabular}
\end{table}

\begin{table}[H]
  \centering
  \caption{ECF net wins over the pre-ECF baseline, with no regression.}
  \label{tab:content-fusion-audit}
  \scriptsize
  \setlength{\tabcolsep}{3pt}
  \renewcommand{\arraystretch}{1.10}
  \begin{tabular}{lrrrr}
    \toprule
    Setting & w/o ECF & with ECF & Net wins & Losses \\
    \midrule
    SWE-bench Verified & 421/500 & \textbf{426/500} & +5 & 0 \\
    SWE-bench Multilingual & 230/300 & \textbf{236/300} & +6 & 0 \\
    Defects4J & 78/371 & \textbf{87/371} & +9 & 0 \\
    \bottomrule
  \end{tabular}
\end{table}

\subsubsection{RQ-2.1: Which components account for the gains?}
Table~\ref{tab:component-ablation} shows a clear division of labor.
Repair-neighborhood fusion is the backbone on Multilingual and Defects4J, the largest single drop there when removed (Table~\ref{tab:component-ablation}).
On Verified, by contrast, the local edit path and its guards matter most, and representative selection with pool-level routing adapts the backbone to this per-pool structure.
ECF sits on top as a sparse content-fusion layer that lifts the final counts by $+5/+6/+9$, with no observed regression (Table~\ref{tab:content-fusion-audit}).

Across the three benchmark settings, ECF makes 20 label-changing corrections.
By construction, where the emitted diff is inspectable on the two SWE-bench pools, ECF's correct outputs are patches it \emph{constructs} rather than copies: the submitted diff is a supported scope-union or contraction absent from the candidate pool in 8 cases (including \texttt{django\_\_django-14315}), while 4 Multilingual outputs re-derive an already-solving candidate's edit.
Of these 12 inspectable correct outputs, 11 flip a previously unresolved instance and form the $+5$ and $+6$ SWE-bench net wins of Table~\ref{tab:content-fusion-audit}, while the 12th re-applies a correct construction on an instance the pre-ECF representative already solved and so leaves the count unchanged rather than adding a win.

ECF's scope localization is not Python-specific: alongside the Python AST locator, which covers $98.0\%$ of Verified patches, tree-sitter grammars resolve a real AST method or class scope for $94.1\%$ of the non-Python Multilingual patches (Java, PHP, and Ruby at $100\%$; Rust $95.0\%$; Go $92.5\%$; JS/TS $82.8\%$; C/C++ $82.3\%$), leaving a brace-balanced block only as a residual fallback.
The small Multilingual takeover count ($9$, all resolved) is therefore bounded by ECF's conservative support and compactness gates, not by coarse scoping.

\subsubsection{RQ-2.2: What does the pipeline still miss?}
We inspect pre-ECF candidate-reachable misses, bugs whose pool contains a solved patch while the routed pre-ECF representative is unsolved.
Same-neighborhood representative choice dominates: it covers 42/55 of these misses (16/22 Verified, 26/33 Multilingual); the remaining 13 are isolated singletons, with none in a different non-singleton neighborhood, so the residual is not a missing repair surface but the wrong close variant chosen within it.

\begin{table}[h]
  \caption{Static relations in pre-ECF same-neighborhood misses.}
  \label{tab:same-neighborhood-gap-taxonomy}
  \centering
  \begin{tabular}{lrr}
    \toprule
    Static relation to solved patch & Multi. & Verified \\
    \midrule
    Same files & 23 & 12 \\
    Higher exact support & 2 & 1 \\
    Higher signature support & 2 & 1 \\
    Solved shorter & 15 & 9 \\
    Solved longer & 19 & 9 \\
    More supported atoms & 11 & 8 \\
    Fewer unsupported atoms & 7 & 10 \\
    Lower token centrality & 26 & 12 \\
    Selected dominates & 11 & 5 \\
    \bottomrule
  \end{tabular}
\end{table}

Table~\ref{tab:same-neighborhood-gap-taxonomy} breaks these by pre-ECF static relations: a solved alternative usually shares the file surface and is less token-central than the chosen representative, yet in a large share every static criterion still ranks the representative ahead of it.
The next distinguishing signal is therefore edit necessity rather than another whole-diff score.

\findx{\findtag{[RQ-2] Findings.} (1) ECF makes 20 label-changing corrections with no regression, 8 by constructing a patch that appears in no candidate.
(2) The remaining gap is not finding the repair but the choice within it: 42 of 55 residual misses pick the wrong close variant in the right neighborhood.
\findtag{Insight.} The last gap is not coverage but a within-neighborhood choice, a single signal (\emph{edit necessity}) that no whole-diff score can express.}

\subsection{RQ-3: Post-Generation Cost}

\noindent\textbf{[Experimental Design]}:
We place every method from Table~\ref{tab:main-fixed-pool} on a cost-versus-repairs plane: the $x$-axis is post-generation wall-clock per bug from Section~\ref{sec:experimental-setup}, and the $y$-axis is the SWE-bench Verified solved count.
We also read the dollar cost of the model-based rows from cached prompt sizes and note whether each method requires an online model and whether it is deterministic.

\begin{figure}[h]
  \centering
  \footnotesize
  \begin{tikzpicture}
  \begin{axis}[
    width=\columnwidth, height=5.4cm,
    xmode=log, log basis x=10,
    xlabel={Post-generation decision cost (ms/bug, log scale)},
    ylabel={Verified solved (/500)},
    xmin=0.013, xmax=16000,
    ymin=305, ymax=450,
    xtick={0.01,0.1,1,10,100,1000,10000},
    xticklabels={$10^{-2}$,$10^{-1}$,$1$,$10$,$10^2$,$10^3$,$10^4$},
    ytick={320,340,360,380,400,420,440},
    tick label style={font=\scriptsize},
    label style={font=\scriptsize},
    grid=major, grid style={black!8, line width=0.3pt},
    axis line style={black!55},
    clip=false,
  ]
    \addplot[draw=black!45, dashed, line width=0.7pt, forget plot]
      coordinates {(0.013,443)(16000,443)};
    \node[font=\scriptsize, text=black!55, anchor=north west] at (axis cs:0.014,442)
      {candidate-reachable ceiling (443)};

    \addplot[only marks, mark=*, mark size=1.5pt, color=black!42, forget plot]
      coordinates {(0.02,382)(0.03,397)(0.07,384)(0.22,403)(0.25,403)(0.53,400)};
    \addplot[only marks, mark=*, mark size=1.9pt, color=black!72, forget plot] coordinates {(0.02,403)};
    \node[font=\scriptsize, text=black!72, anchor=west] at (axis cs:0.03,405) {token medoid};
    \node[font=\scriptsize, text=black!45, anchor=north, align=center] at (axis cs:0.10,380)
      {cheap deterministic\\selectors};

    \addplot[only marks, mark=*, mark size=1.9pt, color=orange!80!black, forget plot] coordinates {(77.1,390)};
    \node[font=\scriptsize, text=orange!58!black, anchor=south] at (axis cs:77,391.5) {LM naturalness};

    \addplot[only marks, mark=triangle*, mark size=2.6pt, color=purple!72!black, forget plot] coordinates {(1260,396)};
    \node[font=\scriptsize, text=purple!72!black, anchor=south] at (axis cs:1260,397.5) {DeepSeek judge};

    \addplot[only marks, mark=triangle*, mark size=2.6pt, color=red!72!black, forget plot] coordinates {(6024,317)};
    \node[font=\scriptsize, text=red!72!black, anchor=south east] at (axis cs:5400,317) {DeepSeek fusion};

    \addplot[only marks, mark=star, mark size=3.8pt, color=blue!62!black, line width=0.7pt, forget plot] coordinates {(3.28,426)};
    \node[font=\scriptsize\bfseries, text=blue!62!black, anchor=south west] at (axis cs:4.4,424.0) {\method};
  \end{axis}
  \end{tikzpicture}
  \caption{Decision-stage cost versus repairs on SWE-bench Verified.}
  \Description{A scatter plot with a logarithmic x-axis of post-generation cost in milliseconds per bug and a y-axis of SWE-bench Verified solved count out of 500. PatchFusion sits at the top left at 3.28 milliseconds and 426 solved, just below a dashed candidate-reachable ceiling line at 443. A cluster of cheap deterministic selectors, including token medoid at 403, sits at the far left between 382 and 403 solved. The LM naturalness ranker is at 77.1 milliseconds and 390 solved. The DeepSeek listwise judge is at 1260 milliseconds and 396 solved. The DeepSeek free-form fusion is at 6024 milliseconds and only 317 solved, at the bottom right.}
  \label{fig:cost-solved}
\end{figure}
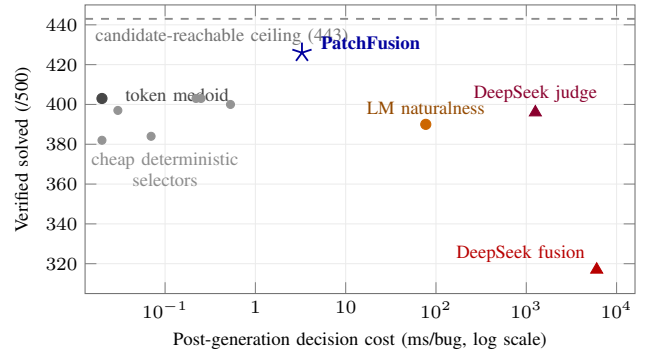

\noindent\textbf{[Experimental Results]}:
Figure~\ref{fig:cost-solved} puts \method in the top-left corner: it solves the most bugs (426/500) at 3.28 ms per bug with no model call, within 17 of the candidate-reachable ceiling.
This 3.28 ms is the decision step in isolation: a median over 7 runs, measured after the candidate pool is loaded and its views are built, and it excludes candidate generation, any model or API call, official-runner execution, and the one-time preprocessing (pool loading, diff normalization, scope building) that precedes the decision; the comparison below is therefore of decision-stage latency, not end-to-end pipeline time, and every method on the plane is timed under the same post-preprocessing convention.
The deterministic consensus and rank-aggregation selectors are equally cheap but capped below it (token medoid strongest at 403), while the model-based methods sit two to three orders of magnitude to the right without overtaking \method, free-form fusion being both the most expensive (6024 ms) and the lowest-scoring (317/500).
The dollar gap is wider still: \method spends no model tokens, whereas each DeepSeek call consumes 4.0k input tokens (and a generated patch for fusion), adding a per-bug API charge and an online-model dependency at a step that should add milliseconds.
\method is also deterministic, returning the same patch from the same pool on every run, while the model-based rows depend on sampling and an external endpoint.

With its decision-time constraints relaxed, a frontier model can be pushed toward a higher plausible-patch yield, but that is a different operating point: it is expensive, nondeterministic, and emits an unauditable program rather than a contraction of a submitted candidate.
Under the fixed-pool decision-time budget studied here, that option is unavailable.

\findx{\findtag{[RQ-3] Findings.} (1) Accuracy does not cost more: \method is the most accurate selector yet decides in 3.28 ms with no model call.
(2) The evaluated model-based rows do not improve accuracy: they cost two to three orders of magnitude more (1260 to 6024 ms vs 3.28 ms) without doing better, and the model-based fusion row is the most expensive (6024 ms) and least accurate (317/500).
\findtag{Insight.} Under this fixed-pool setting there is no speed-accuracy tradeoff: the evaluated model-based selectors do not choose better, so the decision should stay cheap and deterministic.}

\subsection{RQ-4: Robustness to Pool Composition}

\noindent\textbf{[Experimental Design]}:
At every subset size $k$ we evaluate all $\binom{n}{k}$ subsets of each pool and record \method's diff-level solved count as a fraction of that subset's candidate-reachable ceiling, which rescales the three pools onto one axis.
The SWE-bench pools are subsampled by source (Verified $n=6$, Multilingual $n=7$, each source a different system); the Defects4J pool is subsampled by pass@$k$ sample ($n=10$ stochastic samples of one fine-tuned model), a single-model control where any complementarity comes from sampling alone.

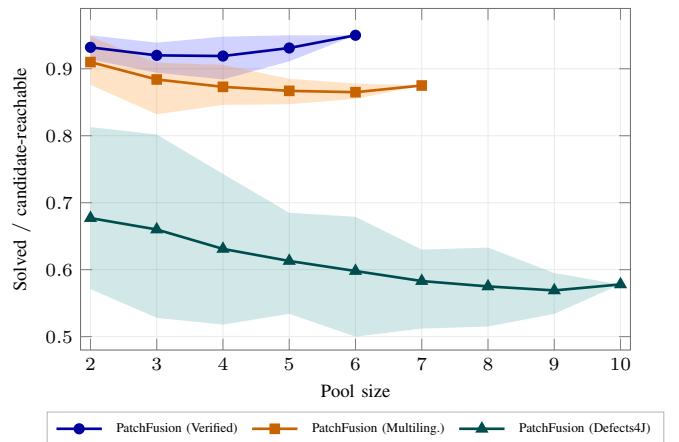
\begin{figure}[h]
  \centering
  \footnotesize
  \begin{tikzpicture}
  \begin{axis}[
    width=\columnwidth, height=6.1cm,
    xlabel={Pool size},
    ylabel={Solved $/$ candidate-reachable},
    xmin=1.85, xmax=10.15, ymin=0.48, ymax=0.99,
    xtick={2,3,4,5,6,7,8,9,10},
    ytick={0.5,0.6,0.7,0.8,0.9},
    tick label style={font=\scriptsize},
    label style={font=\scriptsize},
    legend style={font=\tiny, at={(0.5,-0.18)}, anchor=north, legend columns=3,
      draw=black!28, column sep=3pt},
    grid=major, grid style={black!8, line width=0.3pt},
    axis line style={black!55},
  ]
    \addplot[name path=vmin, draw=none, forget plot]
      coordinates {(2,0.914)(3,0.894)(4,0.884)(5,0.911)(6,0.950)};
    \addplot[name path=vmax, draw=none, forget plot]
      coordinates {(2,0.950)(3,0.939)(4,0.948)(5,0.950)(6,0.950)};
    \addplot[blue, fill opacity=0.18, draw=none, forget plot] fill between[of=vmin and vmax];
    \addplot[blue!62!black, line width=1pt, mark=*, mark size=1.6pt]
      coordinates {(2,0.932)(3,0.920)(4,0.919)(5,0.931)(6,0.950)};
    \addlegendentry{\method (Verified)}

    \addplot[name path=mmin, draw=none, forget plot]
      coordinates {(2,0.876)(3,0.832)(4,0.846)(5,0.847)(6,0.855)(7,0.875)};
    \addplot[name path=mmax, draw=none, forget plot]
      coordinates {(2,0.948)(3,0.909)(4,0.906)(5,0.885)(6,0.878)(7,0.875)};
    \addplot[orange, fill opacity=0.22, draw=none, forget plot] fill between[of=mmin and mmax];
    \addplot[orange!78!black, line width=1pt, mark=square*, mark size=1.5pt]
      coordinates {(2,0.910)(3,0.884)(4,0.873)(5,0.867)(6,0.865)(7,0.875)};
    \addlegendentry{\method (Multiling.)}

    \addplot[name path=dmin, draw=none, forget plot]
      coordinates {(2,0.571)(3,0.528)(4,0.518)(5,0.534)(6,0.500)(7,0.512)(8,0.515)(9,0.534)(10,0.578)};
    \addplot[name path=dmax, draw=none, forget plot]
      coordinates {(2,0.813)(3,0.802)(4,0.743)(5,0.685)(6,0.679)(7,0.630)(8,0.633)(9,0.595)(10,0.578)};
    \addplot[teal, fill opacity=0.18, draw=none, forget plot] fill between[of=dmin and dmax];
    \addplot[teal!60!black, line width=1pt, mark=triangle*, mark size=1.9pt]
      coordinates {(2,0.677)(3,0.660)(4,0.631)(5,0.613)(6,0.598)(7,0.583)(8,0.575)(9,0.569)(10,0.578)};
    \addlegendentry{\method (Defects4J)}
  \end{axis}
  \end{tikzpicture}
  \caption{Fraction of candidate-reachable bugs solved versus pool size.}
  \label{fig:source-subsampling}
\end{figure}

\begin{table}[h]
  \centering
  \caption{Sensitivity of \method.}
  \label{tab:sensitivity}
  \scriptsize
  \setlength{\tabcolsep}{4pt}
  \renewcommand{\arraystretch}{1.15}
  \begin{tabular}{@{}lccl@{}}
    \toprule
    Constant & Default & Swept & Effect \\
    \midrule
    Base-overlap boundary & $0.8$ & $[0.60,0.95]$ & D4J solved $87$ ($85$ at $0.60$) \\
    Atom support & $2$ & $3$ & emitted patches $-1$ \\
    Scope-union cap $K$ & $3$ & $2,\,4$ & emitted patches $+1,\;0$ \\
    Fused-token budget & $160/180$ & $\pm50\%$ & emitted patches $0$ \\
    \bottomrule
  \end{tabular}
\end{table}

\noindent\textbf{[Experimental Results]}:
On the diverse SWE-bench pools \method captures a near-constant high fraction of the reachable bugs at every pool size (91.9 to 95.0\% on Verified and 86.5 to 91.0\% on Multilingual), so its effectiveness is set by the complementarity the pool contains, not by a particular source count.
On the single-model Defects4J pool the captured fraction is much lower and declines as samples accumulate (67.7\% at 2 samples down to 57.8\% at 10), because added samples of one model grow the reachable set faster than the selector keeps pace.

The single-model control isolates the source of that complementarity.
Defects4J is a single-model pass@10 pool, yet \method still gains there: it reaches 87/371 plausible patches, well above the best single sample at 62 and uniform random at 47.3, and ECF lifts the pool from 78 to 87.

\findx{\findtag{[RQ-4] Findings.} (1) Complementarity does not require diverse systems: one model's pass@10 sampling already fuses, lifting the single-model pool from 62 to 87.
(2) The single-model pool's declining captured share (67.7\% to 57.8\%) is a faster-rising reachable ceiling, not fusion degrading; diverse pools stay at a near-constant high share.
\findtag{Insight.} Fusion depends on \emph{complementarity}, not source count: even one model's repeated sampling supplies it, while diverse systems make far more of it reachable.}

\section{Threats to Validity}
\label{sec:threats}

\smallskip\noindent\textbf{Internal validity.}
The main threat is decision-time leakage: a selector could exploit test outcomes, labels, or run metadata rather than the patches themselves.
We bound it with the candidate view of Section~\ref{sec:candidate-interface}, which exposes only diff-derived fields, duplicate counts, and unlabeled pool structure, with labels attached only after the final patch is fixed.
To limit benchmark overfitting, \method's few numeric constants were frozen before evaluation with no development set, and the sensitivity sweep (Table~\ref{tab:sensitivity}) shows that no setting regresses.

\smallskip\noindent\textbf{External validity.}
The candidate pool is a public-leaderboard snapshot, so the results could be specific to it.
We mitigate this by evaluating complete pools, by exceeding the best single source on all three pools (Table~\ref{tab:main-fixed-pool}), and by the RQ-4 complementarity control (Figure~\ref{fig:source-subsampling}), which shows that the gain tracks pool complementarity rather than source count; we therefore scope our strongest captured-fraction results to multi-source pools, while the single-model Defects4J control indicates the gain does not require diverse systems.

\smallskip\noindent\textbf{Construct validity.}
On Defects4J, plausible does not imply correct, so we make the SWE-bench official-runner solved counts the primary result and treat Defects4J as a single-model control reported as a plausible count from MORepair's stored validation.
Every final patch, including the new diffs ECF emits, is scored by each benchmark's official oracle, never a self-reported one.

\section{Related Work}
\label{sec:related}

\subsection{Program Repair, Coding Agents, and Candidate Pools}

Automated program repair (APR) has progressed from search- and heuristic-based generation~\cite{legoues2012genprog,long2016correctcode,wen2018capgen}, through template- and pattern-based fixing~\cite{liu2019tbar,koyuncu2020fixminer} and semantic or constraint-based synthesis~\cite{nguyen2013semfix,mechtaev2016angelix}, to learning-based transformations~\cite{li2020dlfix,lutellier2020coconut,chen2021sequencer}.
LLM coding agents have since expanded the scale and diversity of candidate patches.
Zero-shot repair~\cite{xia2022zeroshot}, pretrained repair models~\cite{fan2023aprllm,xia2023llmapr}, code-completion repair~\cite{li2023cctest}, conversational repair~\cite{xia2024conversation,yang2024morepair}, retrieval-augmented generation~\cite{wang2023rapgen}, repository-aware repair~\cite{yang2025repokg}, agentless pipelines~\cite{xia2025agentless}, and software agents~\cite{yang2024sweagent,zhang2024autocoderover,bouzenia2025repairagent} all improve how candidate patches are produced.
These systems define the upstream generators and trajectories from which a candidate pool is obtained.
As generators improve, pass@$k$ increasingly exposes reachable repairs, but deployment still needs one auditable patch before validation or review.

\subsection{Patch Selection, Ranking, and Fusion}

Patch correctness assessment and ranking use behavioral or plausibility signals~\cite{qi2015plausibility,xiong2018patchcorrectness}, dynamic impact~\cite{ghanbari2022impact}, failing-test similarity~\cite{tian2022failingtests}, and naturalness~\cite{kang2022lmpatch}.
They also use LLM judging~\cite{zhou2024llmapca}, similarity- or history-based reranking~\cite{ghanbari2022similarityranking}, learned patch representations~\cite{ye2021patchassessment,tang2024patchrepr}, and rank aggregation~\cite{dwork2001rankaggregation}.
\method differs by fusing evidence from exact and near-duplicate variants within a repair neighborhood before selecting an auditable representative.

Multi-sample code generation also uses consistency and clustering: AlphaCode clusters generated programs~\cite{li2022alphacode}, self-consistency marginalizes sampled reasoning paths~\cite{wang2023selfconsistency}, CodeT selects code with generated tests~\cite{chen2023codet}, and Universal Self-Consistency uses an LLM to choose among free-form candidates~\cite{chen2024usc}.
Validation-aware agents such as CodeMonkeys and SWE-Search add generated tests, validation-aware selection, or search \cite{ehrlich2025codemonkeys,antoniades2024swesearch}.
In contrast, \method uses repeated edit atoms for evidence-constrained fusion while keeping the final patch auditable against submitted candidates.

Patch construction and multi-hunk repair systems synthesize or refine patches during generation: multi-edit synthesis~\cite{saha2019hercules,wong2021varfix}, iterative refinement~\cite{ye2023iter}, and large-change construction~\cite{li2024giantrepair,liu2026siblingrepair}.
Repository-scale APR and agent benchmarks then expose fixed multi-run candidate pools~\cite{jimenez2024swebench,chen2024repobugs}.
\method starts after generation and turns that pool into one auditable patch.

\section{Conclusion}

We propose \method, a deterministic approach that fuses a fixed pool of candidate patches into a single auditable final patch via repair-neighborhood fusion, representative selection, and evidence-constrained fusion (ECF).
We evaluate \method on \dataset, covering SWE-bench Verified, SWE-bench Multilingual, and Defects4J, where it solves 426/500 and 236/300 SWE-bench bugs and reaches 87/371 plausible Defects4J patches, outperforming consensus and LLM-based selectors at a 3.28 ms per-bug decision with no model cost.
Ablations show that ECF is a content-fusion layer that recovers mis-selected repairs by constructing patches absent from the pool, with no observed regression. Future work can investigate combining our deterministic, test-free patch fusion with an LLM-based agentic workflow to recover more diverse and accurate repairs.


\bibliographystyle{IEEEtran}
\bibliography{references}

\end{document}